\documentclass[twocolumn]{emulateapj}

\newcommand{\xd}{\chi^2/dof}

\newcommand{\Msun}{\hbox{$\rm\thinspace M_{\odot}$}}

\slugcomment{}
\shorttitle{Truncated Accretion Disk of NGC 4593}
\shortauthors{Markowitz \& Reeves}
\begin{document}
\title{A {\it Suzaku} Observation of NGC 4593: Illuminating the Truncated Disk }
\author{A.G.\ Markowitz}
\affil{Center for Astrophysics and Space Sciences, University of California, San Diego, M.C.\ 0424, La Jolla, CA, 92093-0424, USA}

\author{J.N.\ Reeves}
\affil{Astrophysics Group, School of Physical and Geographical Sciences, Keele University, Keele, Staffordshire, ST5 5BG, UK}
\begin{abstract}

We report results from a 2007 {\it Suzaku} observation of the Seyfert 1 AGN NGC 4593.
The narrow Fe K$\alpha$ emission line has a 
FWHM width $\sim$ 4000 km s$^{-1}$, indicating emission from $\gtrsim$ 5000 $R_{\rm g}$.
There is no evidence for a relativistically broadened Fe K line, consistent with the presence of a radiatively efficient 
outer disk which is truncated or transitions to an 
interior radiatively inefficient flow.

The {\it Suzaku} observation caught the source in a low-flux state; 
comparison to a 2002 {\it XMM-Newton} observation indicates that the hard X-ray flux decreased 
by 3.6, while the Fe K$\alpha$ line intensity and width $\sigma$ each
roughly halved. 
Two model-dependent explanations for the changes in Fe K$\alpha$ line profile are explored.
In one, the Fe K$\alpha$ line width has decreased from $\sim$10000 to $\sim$4000 km s$^{-1}$ from 2002 to 2007, suggesting that the thin
disk truncation/transition radius has increased from 1000--2000 to $\gtrsim$5000 $R_{\rm g}$.
However, there are indications from other compact accreting systems that 
such truncation radii tend to be associated only with 
accretion rates relative to Eddington much lower than that of NGC 4593.
In the second model, the line profile in the {\it XMM-Newton} observation consists of a time-invariant
narrow component plus a broad component originating from the inner part of the truncated disk
($\sim$300 $R_{\rm g}$) which has responded to the drop in continuum flux.
The Compton reflection component strength $R$ is $\sim$ 1.1, 
consistent with the measured Fe K$\alpha$ line total equivalent width 
with an Fe abundance 1.7 times the solar value. 
The modest soft excess, modeled well by either thermal bremsstrahlung emission or by
Comptonization of soft seed photons in an optical thin plasma, 
has fallen by a factor of $\sim$20 from 2002 to 2007, 
ruling out emission from a region 5 lt-yr in size.

\end{abstract}

\keywords{galaxies: active --- galaxies: Seyfert --- X-rays: galaxies --- galaxies: individual (NGC 4593) }

\section{Introduction}


Accretion onto supermassive black holes in many Active Galactic Nuclei (AGN)
is generally thought to proceed via a radiatively efficient, optically thick, geometrically thin
disk (``$\alpha$-disk'', e.g., Shakura \& Sunyaev 1973), as evidenced by their optical/UV continua
(the ``big blue bump''; Sun \& Malkan 1989). Galactic black hole systems
also contain evidence for such a component; the thermal blackbody emission extends into the soft X-ray band.
In Seyferts, the Fe K$\alpha$ emission line at 6.4 keV line is a key tracer of the radiatively efficient circumnuclear material. 
Narrow lines with Doppler-broadened FWHMs of a few thousand km s$^{-1}$ are virtually ubiquitous in X-ray spectra of low-redshift 
Seyferts observed with {\it XMM-Newton}, {\it Chandra}-HETGS or {\it Suzaku} (e.g., Yaqoob \& Padmanabhan 2004, Nandra 2006)
and indicate material originating light-days or farther from the black hole, e.g., in the outer disk or molecular torus.
Fe K$\alpha$ line emission originating near the innermost stable orbit of the disk yields a broad 
(FWHMs $\sim$ 0.1$c$) and redshifted profile sculpted by general and special relativistic effects in the regime 
of strong gravity (Fabian et al.\ 1989, 2002),  
although to accurately gauge the strength of the line,
one must correctly model any 
continuum curvature which may be associated with ionized absorption along the line of sight
(e.g., Reeves et al.\ 2004, Turner et al.\ 2005).

However, the flow in compact systems accreting at relatively low values relative to the
Eddington limit may contain a radiatively inefficient component (a radiatively inefficient accretion flow or RIAF; 
Narayan et al.\ 1998, Quataert 2001); a configuration consisting of an inner RIAF flow
which is surrounded by an $\alpha$-disk beyond some transition radius (e.g., Esin et al.\ 1997)
has been suggested for some low-luminosity AGN (Lu \& Wang 2000 and references therein).
In the case of the Seyfert 1 AGN NGC 4593, modeling of the optical/UV continuum indicates
blackbody emission from a truncated thin disk, with an inner radius of emission 
of 30 $R_{\rm g}$\footnote{1  $R_{\rm g} \equiv GM_{\rm BH}/c^2$} (Lu \& Wang 2000).
An observation with {\it XMM-Newton} in 2002 indicated a FWHM line width of 
11000$\pm$1000 km s$^{-1}$; assuming a black hole mass $M_{\rm BH}$ of $10^{6.8}$ $\Msun$,
this width suggests an inner extent of no less than 
roughly 1000 $R_{\rm g}$ (Brenneman et al.\ 2007, hereafter B07). No broad Fe K$\alpha$ line has been confirmed with 
{\it XMM-Newton} (Reynolds et al.\ 2004, B07).
These results were consistent with those obtained from a {\it Chandra}-HETGS observation 
in 2001, from which Yaqoob \& Padmanabhan (2004)
measured a FWHM line width of $2140^{+13780}_{-1830}$ km s$^{-1}$.
An earlier {\it ASCA} observation in 1994 indicated line emission from 
a radius of $\geq$30$R_{\rm g}$, consistent with 
the thin disk truncation radius suggested from optical/UV continuum (``big blue bump'')
spectral energy distribution (SED) fitting (Lu \& Wang 2000).
Furthermore, the 2002 {\it XMM-Newton} observation revealed, in addition to the Fe K$\alpha$ core,
line emission from ionized Fe, likely \ion{Fe}{26} (B07). Such line features are relatively rare in Seyfert
X-ray spectra, and 
could potentially originate in the collisionally ionized 
transition region between the inner radiatively inefficient flow and the thin
disk (e.g., Lu \& Wang 2000). 

Another diagnostic reflection component present in hard X-ray spectra is the
Compton reflection hump peaking at $\sim$20--30 keV, expected when
the hard X-ray power-law continuum illuminates optically thick material.
The strength of the Compton reflection hump $R$ 
was found to be $\sim$ 1 in a 1997 observation with {\it BeppoSAX} (Guainazzi et al.\ 1999);
$R$ = 1 corresponds to a slab covering 2$\pi$ sr of the sky as seen from the illuminating X-ray source.
The {\it Suzaku} X-ray observatory is the first since {\it BeppoSAX}
to provide continuous coverage from below 1 keV to at least 50 keV,
allowing users to spectrally deconvolve the broadband
continuum components (absorbing components, the
primary power-law, and the Compton reflection hump),
but {\it Suzaku}'s lower $>$10 keV background 
yields a more accurate determination of $R$.

In this paper, we report results from an observation of the
nucleus of NGC 4593 with {\it Suzaku} in 2007,
with the goals of constraining the Fe K emission profile and
accurately determining the strength of the Compton reflection hump in order to
constrain the geometry of the circumnuclear accreting gas.
As demonstrated below, {\it Suzaku} caught the source at an atypically low
2--10 keV flux level, a factor of almost 4 lower than 
during the {\it XMM-Newton} observation. 
We observe significant changes in the Fe K$\alpha$ profile
between the 2002 {\it XMM-Newton} and 2007 {\it Suzaku} observations
which may be related to the decrease in continuum flux.
We also report evolution in the strength of the
soft excess, and, tentatively, the ionized Fe K emission.     

The rest of this paper is organized as follows:
Section 2 describes the observations and data reduction.
In Section 3, we present fits to the Fe K emission complex
observed with {\it Suzaku} and compare the results to those
for the 2002 {\it XMM-Newton} observation.
In Section 4, we present fits to the 0.3--76 keV broadband
{\it Suzaku} time-averaged spectrum, and again compare the
results to those obtained from {\it XMM-Newton} to investigate
long-term spectral variability of the broadband emission components.
The results are discussed in Section 5.

\section{Observations and Data Reduction}

NGC 4593 ($z$ = 0.00900, Strauss et al.\ 1992; 
$\alpha$ = 12h39m39.4s, $\delta$ = --5d20m39s, J2000.0 coordinates from the
NASA Extragalactic Database) was observed with {\it Suzaku} from 
2007 December 15 at 04:44 UT until 2007 December 17 at 23:20 UT
(observation ID 702040010). We used data from both the
X-ray Imaging Spectrometer (XIS; Koyama et al.\ 2007) CCDs
and the Hard X-ray Detector (HXD; Takahashi et al.\ 2007).
For the HXD, we used only data collected with the PIN
diodes; we did not consider GSO data in the paper due to the
faintness of the source relative to the GSO background.
The XIS-nominal pointing position was used.  
Further details of the {\it Suzaku} observatory are given in
Mitsuda et al.\ (2007).

\subsection{XIS Reduction}

The XIS data used in this paper were version 2.1.6.16 of the screened 
data provided by the {\it Suzaku} team; data
collected within 436 s of passage through the South Atlantic 
Anomaly (SAA) were discarded, and data were selected to be 
at least 5$\arcdeg$ in elevation above the Earth rim 
(20$\arcdeg$ above the day-Earth 
rim).  All XIS data were taken in normal clocking mode.

The XIS consists of four CCDs, numbered 0--3. XIS-0, 2, and 3
are front-illuminated (FI), the fourth (XIS-1) is back-illuminated 
(BI) and features an enhanced soft X-ray response. Use of XIS-2 was discontinued after 2006 November (when it was damaged due to a likely micro-meteoroid impact).
The {\sc cleansis} script was used to remove hot or flickering pixels.
Source spectra were extracted from a 3$\arcmin$ radius centered on the 
source. The background was extracted using four circles of radius 
1.5$\arcmin$, each located $\sim$6$\arcmin$ from the source. 
The XIS-FI CCDs were in 3$\times$3 and 5$\times$5 editmodes, for a net exposure 
time after screening of 118.8 ks per XIS. XIS-1 was also in 3$\times$3 
and 5$\times$5 editmodes, for a net exposure of 118.4 ks. 
Response matrices and ancillary response files (ARFs) were generated for each 
XIS independently using {\sc xissimrmfgen} version 2007-05-14 and {\sc xissimarfgen} version 2008-04-05 (Ishisaki et al.\ 2007). 
The XIS 0 and 3 source and background 
spectra were added using {\sc mathpha};
response files and ARFs were added using {\sc addrmf} 
and {\sc addarf}, respectively.
All spectra were binned to a minimum of 50 counts 
bin$^{-1}$ to allow use of the $\chi^2$ statistic (Gehrels 1986).

To examine the accuracy of the XIS RMFs and determine residual line width due 
e.g., to imperfect CTI correction, we generated spectra for the 
emission lines generated by the $^{55}$Fe 
calibration sources illuminated two corners of each XIS.
Using the above response matrices and ARFs, 
and {\sc XSPEC} v.11.3.2ag, we fit the spectrum for
each calibration source with a model consisting of three Gaussian components
to model emission from Mn K$\alpha_1$, K$\alpha_2$, and  K$\beta$  
(expected line centroids of 5.899, 5.888, and 6.490 keV,
respectively).  The Mn K$\alpha_1$ line energy centroids for the co-added
XIS-FI, XIS-0, XIS-3, and for the XIS-BI were 
5.908 $\pm$ 0.001,
5.906 $\pm$ 0.002,
5.910 $\pm$ 0.002,
and 5.902 $\pm$ 0.003 keV, 
respectively; such discrepancies are consistent with 
the accuracy ($\sim$0.2$\%$ at 
the Mn K$\alpha$ energy) of the energy calibration of the XIS. 
Fitting the co-added XIS-FI calibration source 
lines without the response file, we determined the 
FWHM energy resolution during the observation to be 168 eV.

The 0.4--10 keV light curve, combined from all three operating XISes
and binned to 5760 s, is plotted in the top panel of Fig.\ 1; the variability
amplitude $F_{\rm var}$ (see Vaughan et al. 2003 for definition) is $16.8 \pm 0.8 \%$.

\subsection{HXD-PIN Reduction}

The HXD-PIN is a non-imaging instrument with a 34$\arcmin$ square FWHM field of view.
12--76 keV PIN source spectra were extracted 
from version 2.1.6.16 event files provided by the HXD 
instrument team. PIN background count rates are variable and strongly depend on 
the time since SAA passage (Kokubun et al.\ 2007), data were selected according 
to the following criteria: at least 500 s since SAA passage, geomagnetic
cutoff rigidity (COR) $\geq$ 8 GV,
and day- and night-Earth elevation angles each $\geq$5$\arcdeg$. Instrumental 
(non-X-ray) background spectra for the PIN were provided by the HXD Team, who
generated the background using the calibrated GSO data for the background monitor
("tuned" background, with METHOD=LCFITDT).
The current accuracy of the PIN non-X-ray background (NXB) model 
for a 1 day observation is $\lesssim$1.5$\%$ (1$\sigma$ peak-to-peak residuals) below $\sim$50 keV
(Fukazawa  et al.\ 2009).
Both the source and background spectra were generated with identical good time 
intervals, and the source exposure was corrected for instrument dead time 
(a 6.8$\%$ effect; the background event files were already deadtime-corrected by the
HXD Team). This yielded a good exposure time of 90.2 ks.
To model the contribution to the total background from 
the Cosmic X-ray Background (CXB), a spectrum of the form 
9.0 $\times$ 10$^{-9}$($E$/3keV)$^{-0.29}$ exp($-E$/40keV) erg cm$^{-2}$ 
s$^{-1}$ sr$^{-1}$ keV$^{-1}$ (Gruber et al.\ 1999) was used
(see the Suzaku ABC Guide for further details). 
The total (X-ray plus particle) background 
12--76 keV flux was 8.6 $\times$ 10$^{-10}$ erg cm$^{-2}$ s$^{-1}$.

The source spectrum was binned to a minimum of 400 counts bin$^{-1}$. We used the 
response file ae$\_$hxd$\_$pinxinome4$\_$20080129.rsp. The mean 12--76 keV 
net source flux and count rate were 4.9 $\times$ 10$^{-11}$ erg cm$^{-2}$ s$^{-1}$ 
and 0.077 ct s$^{-1}$, respectively. The 12--76 keV net source light curve, binned to
11520 s (two satellite orbits), is plotted in the bottom panel of Fig.\ 1; error bars include
a $\sim$2$\%$ systematic uncertainty (Fukazawa et al.\ 2009).
The variability amplitude $F_{\rm var}$ was $<$ 13.8$\%$.

There is a possible contaminating X-ray source, 1WGA J1239.7--0526, located about 5$\arcmin$ south of
NGC 4593, and thus in the field of view of the PIN. 
However, as demonstrated in the Appendix, we estimate that
in the 12--76 keV band, this source contributes only 7$\times$10$^{-13}$ erg cm$^{-2}$ s$^{-1}$ 
to the observed 12--76 keV spectrum of NGC 4593 (about 1.5$\%$ of the net source
flux), and can be safely ignored. 

\section{Model Fits to the Fe K Bandpass Spectrum}

All spectral fitting in this paper used {\sc XSPEC} v.11.3.2ag. 
The abundances of Lodders (2003) were used.  
In all fits below, a neutral Galactic column of 1.89 $\times$ 10$^{20}$ cm$^{-2}$ was included (Kalberla et al.\ 2005).
Unless otherwise stated, all errors below correspond to $\Delta$$\chi^2$ = 2.71
(90$\%$ confidence level for one interesting parameter when the errors are symmetric)
with the XIS-BI/XIS-FI normalization left free (best-fit values were usually close to unity)
and the PIN/XIS normalization set at 1.16 (Maeda et al.\ 2008). 

Previous soft X-ray observations of NGC 4593 have indicated absorption features
attributed to an outflowing, warm absorber system.
Using data from a 108 ks {\it Chandra}-LETGS observation in 2001 February 
and a 10 ks {\it XMM-Newton} observation in 2000 July, Steenbrugge et al.\ (2003)
modeled two zones of ionized absorption along the line of sight:
a highly-ionized zone with log$\xi$ = 2.61$\pm$0.09 erg cm s$^{-1}$ and a column density 
$N_{\rm H} = 1.6 \pm 0.4 \times 10^{21}$ cm$^{-2}$, plus a more lowly-ionized
zone with ionization parameter log$\xi$ near 0.5 erg cm s$^{-1}$ and $N_{\rm H}$ near $6 \times 10^{19}$ 
cm$^{-2}$; 
$\xi \equiv L_{\rm ion} n_{\rm e}^{-1} r^{-2}$, where $L_{\rm ion}$ is defined as
the isotropic 1--1000 Ryd ionizing continuum luminosity, $n_{\rm e}$ is the electron number density, and $r$ 
is the distance from the central continuum source to the absorbing gas.
Both zones were measured to be outflowing at --400 km s$^{-1}$ relative to systemic.
From an 80 ks observation with {\it Chandra}-HETGS in 2001 June,
McKernan et al.\ (2003) confirmed the highly-ionized absorber, deriving
$N_{\rm H} = 5.37^{+1.45}_{-0.79} \times 10^{21}$ cm$^{-2}$ and  log$\xi$ = $2.52^{+0.06}_{-0.04}$ erg cm s$^{-1}$,
although they derived a lower outflow velocity relative to systemic.    
The high-column, high-ionization component manifests mainly itself via  
H- and H-like lines, the most prominent ones being
\ion{N}{7}, \ion{O}{8}, \ion{Ne}{9}, \ion{Ne}{10}, \ion{Mg}{12}, \ion{Si}{13}, and \ion{Si}{14},
as well as an absorption edge due to \ion{O}{8},
There is also line absorption due to Fe L {\sc XX--XXV} which, in CCD resolution spectra,  
yields an absorption trough near 0.9 keV and a series of blended edges
which yield a trough from $\sim$1.3 to 2.2 keV.
The low-column, low-ionization component  manifests itself via \ion{O}{5} and \ion{O}{6} absorption
and is not expected to contribute strongly to CCD resolution spectra above 0.2 keV.
McKernan et al.\ (2003) also reported 
a neutral Fe L$_3$ edge at 707 eV  due to dust along the line of sight in the host galaxy of NGC 4593 
(this feature was also detected by B07 in their analysis of the {\it XMM-Newton} pn spectrum).
A moderately strong, steep soft excess has been known since an {\it EXOSAT} observation by 
Pounds \& Turner (1988) and has been phenomenologically modeled
using thermal bremsstrahlung and Comptonization models (e.g., B07). 

To fit the Fe K bandpass, we restricted ourselves to the $>$4--11.5\,keV band of the co-added XIS-FI spectrum,
where warm and neutral absorption and the soft excess are not expected to have an impact.
Data/model residuals to a simple power-law are plotted in Figure 2a, 
and reveal that in addition to the clear Fe K$\alpha$ emission line at 6.4 keV,
there is emission near 7.0 keV (all photon energies are rest frame unless
otherwise stated). A model consisting of a power-law plus a Gaussian component
to model the Fe K$\alpha$ line yields $\xd$=360.8/318; residuals are plotted in Fig.\ 2b. 
We added a Gaussian with energy centroid fixed at 7.056 keV  to model Fe K$\beta$ emission,
with the normalization held at 0.13 times that of the Fe K$\alpha$ emission line (i.e., assuming
K$\beta$ emission from neutral Fe; untying this parameter did not yield a significant improvement
in this or any subsequent fit). $\xd$ fell to 332.3/318.  
Data/model residuals are plotted in Fig.\ 2c.  To model the effects of an Fe K edge at 7.11 keV due to Compton reflection,
we added a {\sc pexrav} component (Magdziarz \& Zdziarski 1995), 
with an inclination angle of 30$\degr$ assumed,
and a power-law cutoff set at 500 keV. The reflection strength $R$ was kept fixed at 1.08
and the Fe abundance was kept at solar,
as per the best-fit model to the 0.3--76 keV {\it Suzaku} spectrum (see Section 4).
$\xd$ fell to 320.6/318. 
This is our best-fit model to the Fe K bandpass spectrum; best-fit parameters are given in Table 1.

Data/model residuals are plotted in Fig.\ 2d. There appear to be some moderate ($\sim$7$\%$)
positive residuals at 6.7 keV, the energy of \ion{Fe}{25}. However, there was
no significant improvement to the fit when we added another
Gaussian component with energy centroid fixed at 6.70 keV and width $\sigma$ fixed at 1 eV.
$\chi^2$ fell by only 2.5 for one less $dof$, significant at only the 88$\%$ confidence 
level according to an $F$-test (and at 87.5$\%$ according to Monte Carlo 
simulations.)\footnote{As pointed out by Protassov et al.\ (2002), it is improper to use the $F$-test to determine the
statistical significance associated with comparing a model which contains an emission or absorption line
to a ``null hypothesis'' model where the line does not exist.  This is because the $F$-test cannot be used when
the null values of one parameter lie on the boundary of possible parameter values.
In the case of any emission or absorption line, the boundary condition of zero line flux
occurs for the null hypothesis model.
When the $F$-test is used in this manner, the resulting statistical significance is at best only 
an ``estimate'' of the true stastical significance. 
In this paper, we include the results from the $F$-test, but we also include results using 
Monte Carlo simulations, described in detail in Section 3.3 of Porquet et al.\ (2004).
In these simulations, one 
searches over the possible energy ranges where a feature may be expected to be detected,
fitting an unresolved Gaussian at multiple energy bins to test if fitting spurious features due to photon noise can yield 
a reduction in $\chi^2$ as large as the reduction associated with
including the emission/absorption line in the model.
In the case of a line with an expected observed energy, e.g., 6.70 keV in the case
of an expected \ion{Fe}{25} emission line, we searched over an energy range which was centered on the
line expected energy and had a width $\pm$ the FWHM resolution of the instrument.
Throughout this paper, we list the statistical significances determined from
Monte Carlo simulations in parentheses following the significances estimated using
the $F$-test.}
The upper limit on the intensity (equivalent width, $EW$) to \ion{Fe}{25} emission
was $2.8 \times 10^{-6}$ ph cm$^{-2}$ s$^{-1}$ (19 eV).

B07 reported detection of an emission line near 6.9 keV, likely associated with
\ion{Fe}{26}. We added to our model a narrow (width $\sigma$=1 eV)
Gaussian component with energy centroid fixed at 6.96 keV, but no there was no significant
improvement to the fit, as $\vert\Delta\chi^2\vert$ was $<$ 2.71 for one less $dof$. 
The upper limit on the intensity ($EW$) to \ion{Fe}{26} emission
was $1.8 \times 10^{-6}$ ph cm$^{-2}$ s$^{-1}$ (17 eV).

The Fe K$\alpha$ line's width $\sigma$ is 41$^{+12}_{-15}$  eV
(41$^{+12}_{-16}$ eV after subtracting in quadrature the instrumental broadening of $<$9 eV
based on the fits to the $^{55}$Fe calibration sources in Section 2.1).
The line $EW$ is 255$\pm$19 eV. 
The best-fit energy centroid of the Fe K$\alpha$ line in the
co-added spectrum is 6.421$\pm$0.007 keV, or 6.412$\pm$0.007 keV
after accounting for the small gain offset as indicated by the $^{55}$Fe calibration line.
The best-fit line energies in separate fits to the individual XIS-0 and XIS-3, and XIS-1 spectra
are 6.422$\pm$0.009, 6.408$\pm$0.009 keV, and 6.401$\pm$0.014 keV, respectively, after accounting for the gain offsets 
in each of those detectors. 
The small implied offset from 6.40 keV in the co-added spectrum is likely not real, given that
XIS-3 and XIS-1 each indicate emission consistent with neutral Fe, 
and there may be a slight energy scale problem
associated with XIS 0 under investigation as of this writing.
We also note that the Fe K$\alpha$ 
line width in each XIS is consistent with that from the co-added spectrum. 
 
We tested for the presence of a Compton shoulder,
expected if the bulk of the narrow Fe K$\alpha$ line (hereafter also
referred to as the ``core'') originates in 
Compton-thick material. We modeled such a component using a Gaussian component
with energy centroid fixed at 6.32 keV and width $\sigma$ left 
free.\footnote{The sum of first scatterings of Fe K$\alpha$ photons 
forms a ``shelf''-like feature extending redward from 6.400 keV, with
the maximum downward energy shift due to a single Compton scattering being 0.156 keV; 
the shape of the ``shelf''  and the intensity relative to that of the Fe K$\alpha$ core
are dependent upon geometry, column density of the reflecting material, and
viewing angle (see e.g., Murphy \& Yaqoob 2009; Watanabe et al.\ 2003; Matt 2002).
Given the quality of the current data set and the energy resolution of the XIS,
a Gaussian emission component can be an adequate substitute 
for a first-scattering Compton shoulder
if the energy centroid is constrained to $>$6.244 keV (ideally, 
centered near 6.32 keV) and the width $\sigma$ is $\sim$0.1--0.2 keV.}
We found an upper limit to the
intensity ($EW$) to Compton shoulder emission of $7 \times 10^{-6}$ ph cm$^{-2}$ s$^{-1}$ (44 eV),
or about 24$\%$ that of the Fe K$\alpha$ core.

\subsection{Comparison with the 2002 {\it XMM-Newton} Observation}

To investigate possible evolution of the Fe K emission complex, we
compared the results obtained with {\it Suzaku} to those obtained in
a re-analysis of the 76 ks observation in 2002 June by the {\it XMM-Newton} (ObsID 0059830101).
We downloaded Observation Data Files (ODF) events data and
reprocessed events for the European Imaging Photon Camera (EPIC) pn detector
using XMM Science Analysis Software v.7.1.0.
Further details of the observation are given in B07.
Source and background spectra were extracting using standard analysis methods
for a good exposure time of 53 ks (the pn had been in small window mode to prevent pile-up).
The source spectrum was grouped to a minimum of 40 ct bin$^{-1}$.

We fit the 4--11.5 keV pn spectrum with a model consisting of
a power-law component plus Gaussian components 
to model emission from \ion{Fe}{1} K$\alpha$ and K$\beta$, 
and \ion{Fe}{26}.  We also included a Compton reflection component using {\sc pexrav} with
$R$ fixed at 
1.08.\footnote{Fixing $R$ at 1.08 follows the assumption that the Compton reflection component
has tracked the continuum flux over 5 years.
If the absolute normalization of the Compton hump 
tracked the intensity of the Fe K$\alpha$ line from 2002 to 2007, 
then we would expect $R$
to increase by a factor of 2.24--2.42 (consistent with 
tracking the $EW$ of the Fe line).
In the 2002/{\it XMM-Newton} spectrum, if $R$ 
were really 2.42, then we would observe a significant Fe K edge at 7.1 keV. 
With the lack of $>$ 10 keV coverage, the value of $R$ is poorly constrained, but 
fixing $R$ at 2.42 (or any value $\gtrsim$ 1), 
with solar Fe abundance, yielded edge-like residuals in emission above 7.1 keV.
Thawing the Fe abundance did not yield data/model residuals as good as when $R$ 
had been fixed at 1.08. The exact choice of $R$ in the fits is thus somewhat
arbitrary, but does not significantly affect our conclusions
in the paper regarding the evolution in the observed width of the Fe K$\alpha$ line
or in the soft excess.}  It was not significant to include an edge at 7.11 keV (optical depth $\tau$ $<$ 0.03), so the
choice of $R$ is somewhat arbitrary given that there was no 
simultaneous $>$ 10 keV energy coverage. However, the 
depth of the Fe K edge does not significantly affect the conclusions, and so we include
this component for completeness.  The best-fit parameters 
are listed in Table 1 for comparison with the 2007 Suzaku results; 
parameters are consistent with those obtained by B07 for 
the Fe K$\alpha$ and the \ion{Fe}{26} lines.
Fig.\ 3a shows data/model residuals to a model consisting of only a power-law and a Compton reflection
hump for both the {\it XMM-Newton} and {\it Suzaku} Fe K bandpass spectra; Fig.\ 4 shows contour plots of
line intensity versus centroid energy to this model indicating the significance of including an unresolved Gaussian component. 
Fig.\ 3b shows the data/model residuals when the Fe K$\alpha$ and Fe K$\beta$ lines are additionally modeled. 

We also tested for a possible Compton shoulder,
modeled the same way as in Section 3.1; the upper limit to Compton shoulder emission
intensity ($EW$) was $1.1 \times 10^{-5}$ ph cm$^{-2}$ s$^{-1}$ (22 eV), or 23$\%$ that
of the Fe K$\alpha$ core.


Comparing the results from the 2007 {\it Suzaku} and 2002 {\it XMM-Newton}
observations, several things are apparent.
While the 4--10 keV flux has decreased from 
$2.47 \times 10^{-11}$ erg cm$^{-2}$ s$^{-1}$ in 2002 to $0.69 \times 10^{-11}$ erg cm$^{-2}$ s$^{-1}$ in 2007 (a factor of 3.6), 
the total intensity of the Fe K$\alpha$ line 
has decreased from $4.79 \pm 0.48 \times 10^{-5}$ (2002) to $2.93 \pm 0.22 \times 10^{-5}$ ph cm$^{-2}$ s$^{-1}$ 
(2007), a factor of only 1.6. The Fe K$\alpha$ line $EW$ thus has increased from 114 $\pm$11 eV (2002) to 255$\pm$19 eV (2007). 
In this model where the Fe K$\alpha$ line is modeled with a single
Gaussian component, the width has decreased from 87$\pm$17 eV (2002) to
41$^{+12}_{-16}$ eV (2007). The two line widths are inconsistent up to the
$\Delta\chi^2$ = 9.5 ($>$99.7$\%$, or $>$3$\sigma$) confidence level.
Given that the XIS and the pn have similar energy resolution near 6 keV,
the evolution in line width is likely intrinsic to the source.

Finally, there is highly tentative evidence that the ionized emission blueward of 6.4 keV has evolved, too.
The lower limit to the intensity of the \ion{Fe}{26} line in the pn spectrum and the
upper limit to \ion{Fe}{26} emission (assuming an unresolved line in each case)
are just barely consistent at the 90$\%$ confidence level
Figs.\ 3b and 4 hint at a marginal change in line intensity as well.
We caution, however, that, given the energy resolution and the quality of the spectrum,
there is likely blending between the \ion{Fe}{26} and Fe K$\beta$ lines and Fe K edge at 7.11 keV, and
the width of the \ion{Fe}{26} line is not constrained.

\subsection{Dual-Gaussian Fit to the Fe K$\alpha$ Core}

As an alternative, we considered the possibility that the Fe K$\alpha$ core in NGC 4593
consists of {\it two} Gaussian components, one narrow (width $\sigma$ $\sim$ 40 eV) and one relatively broader,
with $\sigma$ $\sim$ 100 eV. The narrow component is assumed to be present in both spectra.
The broad component is assumed to respond to the drop in continuum flux and assumed to
be detected only in the 2002 {\it XMM-Newton} spectrum and to be
too faint to be detected in the 2007 {\it Suzaku} spectrum. 

We first fit the {\it XMM-Newton} spectrum with this ``dual-Gaussian'' model.
The narrow Gaussian component's width $\sigma_{\rm 1}$ was fixed at 41 eV, the best-fit
value from the {\it Suzaku} fit, and the broader Gaussian component's width $\sigma_{\rm 2}$ was left free.
The intensity of the narrow Gaussian, $I_1$, was initially kept fixed at the
best-fit value from the {\it Suzaku} spectrum, $2.93 \times 10^{-5}$ ph cm$^{-2}$ s$^{-1}$, as might be expected if this 
component has not responded to the drop in continuum flux in the {\it Suzaku} spectrum; thawing this parameter
did not yield a significant improvement in the fit and constraints on $I_1$ were poor.
The intensity of the broad line, $I_2$, was left as a free parameter.
Both Gaussian energy centroids were fixed at 6.40 keV.
In the best-fit model, $\xd$ = 1106.3/1092, $\sigma_{\rm 2} = 177^{+84}_{-52}$ eV,
$I_2 = 2.18^{+0.66}_{-0.34} \times 10^{-5}$ ph cm$^{-2}$ s$^{-1}$.
The $EW$s of the narrow and broad components in the model were 64 eV and $47^{+14}_{-7}$ eV, respectively,
i.e., both components comprise roughly equal fractions of the total line flux.

Returning to the {\it Suzaku} spectrum, we added to our best-fit model a Gaussian component with
width fixed at 177 eV. This did not yield a significant improvement in the
fit; the upper limit to the intensity ($EW$) was $1.0 \times 10^{-5}$ ph cm$^{-2}$ s$^{-1}$ (65 eV).

\section{The Broadband {\it Suzaku} Spectrum}

We fit the XIS-FI data from 0.4--11.5 keV,
the XIS-BI data from 0.3--10.5 keV, and the PIN
data from 12--76 keV. 
Due to calibration uncertainties associated with the instrumental Si K edge,
data from 1.75--1.83 keV were ignored in the XIS-FI and from
1.75--1.87 keV in the XIS-BI data.
All the data are plotted in Figure 5a.

Data/model residuals to a simple power-law component absorbed only by the  
Galactic column are shown in Figure 5b and indicate spectral features
reported in previous studies: an absorption feature near 0.9 keV, the moderate soft excess at lower energies,
and an excess$>$ 12 keV which is likely the Compton reflection hump.

Our ``initial guess'' model was based on the best fit to the {\it XMM-Newton} spectrum by B07.
It consisted of the following components:
a power-law; a Compton reflection hump modeled with {\sc pexrav},
with abundances set to solar values, an inclination angle of 30$\degr$ assumed,
and a power-law cutoff set at 500 keV; and a soft excess phenomenologically
parametrized as thermal bremsstrahlung emission, using the model {\sc zbrem} in {\sc xspec}.
We modeled one zone of ionized absorption with an {\sc xstar} v.2.1l table 
which assumed solar abundances, a turbulent velocity of 200 km s$^{-1}$ and an input photon 
spectrum of 2.
The absorber's outflow velocity relative to systemic was kept fixed at
--400 km s$^{-1}$ (Streenbrugge et al.\ 2003)
The best-fit model had $\xd$ = 1285.4/1110; residuals are plotted in Figure 5c.     

We then added components to model features reported in previous observations.
We added narrow (width $\sigma$ = 1 eV) Gaussian components to model line emission from
\ion{O}{7} $f$ (energy centroid fixed at 561 eV) and \ion{Ne}{9} $f$ (energy centroid fixed at 905 eV), features
reported using {\it Chandra}-HETGS by McKernan et al.\ (2003);
$\xd$ dropped to 1279.8/1109 and then 1272.6/1108, respectively.
We next added an Fe L$_3$ edge with energy fixed at 707 eV, reported
by both McKernan et al.\ (2003) and B07; $\xd$ fell to 1262.1/1107.
A final improvement in the fit was achieved by adding a third narrow Gaussian
near 1.33 keV, the energy of He-like Mg, where there still remained $\sim$+3 to +4$\%$ residuals. 
We assumed emission from \ion{Mg}{11} $f$ 
and fixed the Gaussian energy centroid at 1.331 keV; $\xd$ fell to 1254.5/1106.
Data/model residuals are plotted in Figure 5d.
Best-fit model parameters are listed in Table 2.
An additional layer of absorption, either neutral or ionized, was not required in our model.
It was not significant to thaw the Fe abundance of the {\sc pexrav} component from solar values. 

In the best-fit model, the power-law component had $\Gamma$ = $1.656^{+0.018}_{-0.025}$, and 
the warm absorber was modeled using a column density 
$N_{\rm H,WA}$ = $ 2.8 \pm 0.8 \times 10^{21}$ cm$^{-2}$ and an ionization parameter of 
log$\xi$=2.39 $\pm$ 0.17 erg cm s$^{-1}$.
The best-fit value of $R$ was 1.08$\pm$0.20,
and Fig.\ 6 shows a contour plot of $R$ as a function of $\Gamma$. The uncertainties here are statistical only,
and calculated assuming the PIN/XIS-FI normalization is fixed at 
1.16.\footnote{When the PIN/XIS-FI normalization was allowed to vary, it went to a value above 1.5,
which is highly inconsistent with
{\it Suzaku} calibration. The PIN/XIS-FI normalization was thus kept fixed at 1.16 in all fits.}
Assuming $\pm$1.5$\%$ systematics in the absolute background flux, the systematic uncertainty on $R$ is
an additional $\pm$0.15 (not plotted in Fig.\ 6). 

We explored alternate parametrizations of the soft excess.
Modeling the soft excess as a simple power-law yielded a good fit, with $\xd$ =  
1257.3/1106 for best-fit values of  $\Gamma = 3.6 \pm 1.3$ and power-law normalization
at 1 keV of $2.3^{+14.7}_{-1.9} \times 10^{-5}$ ph cm$^{-2}$ s$^{-1}$ keV$^{-1}$. 
A good fit was also achieved 
assuming thermal Comptonization emission, modeled using {\sc comptt} in {\sc xspec}
(Titarchuk 1994). 
A spherical geometry was assumed, as was an input soft photon temperature of 50 eV.
In the best-fit model, $\xd$ was 1255.8/1105 for a plasma temperature of $53^{+57}_{-31}$ keV
and optical depth $<$0.28.         
In the best-fit model for each cases, data/model residuals and the 0.4--1.0 keV flux of the soft excess component 
were virtually identical to those obtained for the best-fit model
assuming thermal bremsstrahlung emission. 
All other model parameters were consistent with
those obtained when thermal bremsstrahlung emission was modeled.

We next tested if emission from a relativistically broadened Fe K line was required.
We used a {\sc laor} component in {\sc xspec} (Laor 1991).
The inner radius, outer radius, disk inclination, and emissivity index
were initially kept fixed at 1.5 $R_{\rm g}$, 400 $R_{\rm g}$, 30$\degr$ and --3, respectively.
Similar to Reynolds et al.\ (2004) for the 2002 {\it XMM-Newton} spectrum,
no significant improvement to the fit was found
even when thawing the inner radius or disk inclination ($\chi^2$ dropped by less than 2). 
The upper limit to the broad Fe K line intensity ($EW$) was
$4 \times 10^{-6}$ ph cm$^{-2}$ s$^{-1}$ (40 eV).   

We next explored if the soft excess and Compton reflection component could be modeled in a self-consistent manner
by assuming reflection from an ionized surface.
We used the table model reflion.mod (Ross \& Fabian 2005), and set the input photon index equal to that
for the primary power-law. Initially, the Fe abundance relative to solar values $Z_{\rm Fe}$ was kept fixed at 1.0.
In the best-fit model, $\xd$ was 1368/1106 for $\xi$ near 1500 erg cm s$^{-1}$,
with poor data/model residuals. The fit was able to model correctly most
of the soft excess flux, but underpredicted the strength of the
Compton reflection hump, even as the best-fit value of $\Gamma$ went to $\sim$1.0.
Allowing $Z_{\rm Fe}$ to vary did not improve the fit.
We next modeled relativistic blurring of the ionized reflected emission with {\sc kdblur}.
We kept the outer radius of blurring fixed at 400 $R_{\rm g}$ and allowed the inner radius
to vary from 1.24 to 200 $R_{\rm g}$. However, no further significant improvement to the fit
was found for any value of the inner radius tested. Ionized reflection models are not discussed further.

\subsection{A Re-analysis of the 2002 {\it XMM-Newton} pn Broadband Spectrum}

In the best-fit model, the observed (absorbed) 2--10 keV flux was $1.03 \times 10^{-11}$ erg cm$^{-2}$ s$^{-1}$, a factor of 
3.8 lower than that reported by B07 for the 2002 {\it XMM-Newton} observation. This is also a factor of 3.8
lower than the long-term average 2--10 keV flux as determined by continuous {\it Rossi X-ray Timing Explorer} 
Proportional Counter Array ({\it RXTE}-PCA)
monitoring from 2004 February to 2007 December\footnote{We downloaded all the public archive data
for NGC 4593 and created a 2--10 keV light curve, binned for each observation, 
standard extraction methods; the reader is referred to e.g., Markowitz et al.\ (2003)
for details on light curve extraction and to Edelson \& Nandra (1999)
for details regarding on PCA background subtraction.} (Figure 7; see also Summons et al., in prep.).

To explore long-term variability of the spectral components in a model-dependent fashion,
we tried to apply our best-fit {\it Suzaku} model
to the 0.2--12 keV 2002 {\it XMM-Newton} pn spectrum,
while allowing as few parameters as possible to change between models to both spectra.
As per Section 3.2, we allowed the Fe K$\alpha$ line energy centroid,
width, and intensity to vary, and we added a narrow Gaussian
component to model \ion{Fe}{26} line emission.
We modeled the warm absorber using the same {\sc xstar} table model as for the {\it Suzaku} spectrum.

We found that, in order to achieve a good fit and good
data/model residuals ($\lesssim\pm2-3\%$, it was necessary 
and sufficient to thaw the power-law normalization,
power-law photon-index, bremsstrahlung normalization and temperature,
Fe L$_3$ edge depth $\tau_{\rm FeL3}$, and warm absorber column density $N_{\rm H,WA}$. 
Best-fit values for these parameters are listed in 
Table 3, alongside the corresponding values from the {\it Suzaku} fit. 
All other parameters were kept frozen at the values listed in Table 2.
The best-fit model had $\xd$=1793.2/1756; 
data/model residuals are plotted in Fig.\ 8.
We found that it was not necessary to include in the model a second, lower-ionization
warm absorber along the line of sight.

As reported by B07, the total absorbed 0.5--10 keV flux during the 2002 {\it XMM-Newton}
observation was $6.74 \times 10^{-11}$ erg cm$^{-2}$ s$^{-1}$, a factor of
4.7 higher than the total absorbed 0.5--10 keV flux during the 2007 {\it Suzaku}
observation. However, the broadband spectral changes between the two spectra cannot be explained
solely by changes in power-law normalization and photon index;
the best-fit model to the pn spectrum in this case had $\xd \sim 4$
and unacceptable data/model residuals, particularly below 2 keV. 
Our model fits indicate that the difference in $N_{\rm H,WA}$ between the two
observations is not highly statistically significant, as the column densities are consistent at the 
$\Delta\chi^2 \sim 6$ confidence level.  
(The variation in $\tau_{\rm FeL3}$ is not likely real, either;
there is some mild degeneracy between $\tau_{\rm FeL3}$ and $N_{\rm H,WA}$, and the values of 
$\tau_{\rm FeL3}$ are consistent at the $\sim$99$\%$ confidence level ($\Delta\chi^2 = 7$).
If this component is associated with dust in the host galaxy along the line of sight at least several pc from the
black hole, then we would not expect any temporal variation in $\tau_{\rm FeL3}$.)
Evolution in the normalization of the soft excess must contribute to the 
change in soft X-ray flux between the two observations. For comparison purposes,
we now focus on the 0.4--2.0 keV unabsorbed flux, $F_{\rm 0.4-2, unabs}$, of this component.

The normalization of the soft excess in the {\it Suzaku} spectrum is not well constrained
due to its faintness. Assuming thermal bremsstrahlung emission, 
we find $F_{\rm 0.4-2, unabs} =  8.7^{+12.2}_{-6.1} \times 10^{-14}$  erg cm$^{-2}$ s$^{-1}$
for this component in the {\it Suzaku} spectrum
(uncertainties are based on the error on the normalization of this component).
The value of $F_{\rm 0.4-2, unabs}$ of this component in the pn spectrum was
$4.72 \pm 0.19 \times 10^{-12}$ erg cm$^{-2}$ s$^{-1}$.
We conclude that, in the context of modeling the soft excess as
thermal bremsstrahlung emission, the normalization has dropped by at least a factor of 20.
Unfolded model spectra for our best-fit models to the pn and {\it Suzaku} spectra
assuming thermal bremsstrahlung emission are shown in Figs.\ 9 and 10, respectively.

In the {\it XMM-Newton} spectrum, we substituted the thermal bremsstrahlung component with a thermal Comptonization component,
again using {\sc COMPTT}, and keeping the geometry parameter, 
input soft photon temperature, and plasma temperature
fixed at the best-fit values found for the {\it Suzaku} spectrum; only the optical depth and
normalization were allowed to vary. 
This yielded a similar good fit, with $\xd$ =  1922.6/1755;  
best-fit values for the free parameters are listed in Table 3.
$F_{\rm 0.4-2, unabs}$ of the Comptonization component was
$1.54^{+0.33}_{-0.13} \times 10^{-11}$ erg cm$^{-2}$ s$^{-1}$ (flux uncertainty is from the uncertainty
on the normalization of this component). This flux 
is a factor of at least 20 higher than
the corresponding flux from the best-fit {\it Suzaku} model,
$8.2^{+54.6}_{-6.7} \times 10^{-14}$ erg cm$^{-2}$ s$^{-1}$.


\section{Discussion and Conclusions}

\subsection{Summary of Observational Results}

We have presented results from a {\it Suzaku} observation of the nucleus of 
the Seyfert AGN NGC 4593 in 2007 December,
and we compare our spectral fits for both the Fe K bandpass and the broadband X-ray spectrum
with those obtained from a 2002 {\it XMM-Newton} EPIC-pn observation.
{\it Suzaku} caught the source at a relatively low X-ray flux level: the 2--10 keV continuum
flux during the {\it Suzaku} observation was a factor of 3.8 lower.

The Fe K$\alpha$ line intensity has dropped by a factor of 1.7, suggesting that
roughly half of the total line flux has responded to the drop in continuum flux.
One of our main results is that the Fe K$\alpha$ line is significantly more narrow 
in the {\it Suzaku} observation. Modeling the line as a single Gaussian, we find that the
width $\sigma$ has dropped from $87 \pm 17$ eV in 2002 to $41^{+12}_{-16}$ eV
in 2007. We also modeled the line using a dual-Gaussian model composed of relatively
narrow and broad lines. The former dominates the {\it Suzaku} profile and is assumed to be
time-invariant; in the {\it XMM-Newton} spectrum, both lines are modeled to have roughly
equal intensity and the broad component has a width $\sigma = 177^{+84}_{-52}$ eV.
There is highly tentative evidence for the \ion{Fe}{26} emission line at 6.96 keV 
to have dropped in intensity from 2002 to 2007, assuming an intrinsically narrow (unresolved) line: 
in the {\it Suzaku} observation, the line is not significantly detected ($EW < 17$ eV).

In our broadband fits to the 0.3--76 keV spectrum, 
the primary power-law component, commonly attributed to
inverse Comptonization of soft seed photons in a hot corona 
(e.g., Haardt et al.\ 1994), was relatively flat, with $\Gamma$ near 1.65. The 
Compton reflection component had a relative strength $R \sim 1.08$.
We also model a modest soft excess using both thermal bremsstrahlung emission
and thermal Comptonization of soft seed photons, similar to B07, and we obtain similar results.
Importantly, we find the soft excess has dropped in flux by a factor of at least $\sim$20 between
the {\it XMM-Newton} and {\it Suzaku} observations. 
We model one zone of absorption along the line of sight,
the previously seen highly-ionized (log$\xi$ $\sim$ 2.5)  zone, with a column density
similar to that obtained by McKernan et al.\ (2003) and Steenbrugge et al.\ (2003).
There is no strong evidence for evolution of the warm absorber
between the two observations.
A relativistically broadened Fe K$\alpha$ line was not detected in the {\it Suzaku} spectrum;
Reynolds et al.\ (2004) demonstrated a similar result in the {\it XMM-Newton} spectrum.


\subsection{Tracing the Truncated Disk with the Fe K$\alpha$ Line}    

We explore two (model-dependent) scenarios to correlate 
the changes in Fe line intensity and profile with the observed drop in continuum flux.
In the model where a single Gaussian was used to describe the Fe K$\alpha$ profile, 
the width $\sigma$ in the 2007 {\it Suzaku} observation
was $41^{+12}_{-16}$ eV, which corresponds to a FWHM velocity $v_{\rm FWHM}$ of $4420^{+1290}_{-1730}$ km s$^{-1}$.
This is roughly commensurate with the optical broad emission line width:
Peterson et al.\ (2004) reported FWHM H$\alpha$ and H$\beta$ line widths of $3399\pm196$ and $3769\pm862$ km s$^{-1}$,
respectively.\footnote{Peterson et al.\ (2004) reported that the continuum-line lag results were poor.
The H$\alpha$ lag was reported as $3.2^{+5.6}_{-4.1}$ lt-days but Peterson et al.\ (2004) recommended
caution. The H$\beta$ lag was reportedly ``completely unreliable.''}
Assuming that the line originates in gas that is in virialized orbit around the black hole, we can
estimate the distance $r$ from the black hole to the line-emitting gas. 
Assuming that the velocity dispersion is related to 
$v_{\rm FWHM}$ as $<$$v^2$$>$ = $\frac{3}{4}v^2_{\rm FWHM}$ (Netzer 1990), we use $G$$M_{\rm BH}$ = $r$$v^2$. 
We use a black hole mass $M_{\rm BH}$ of $6.6 \times 10^6 \Msun$, an
estimate based on the relation between $M_{\rm BH}$ and
stellar velocity dispersion in Seyferts (Nelson et al.\ 2004). 
The best-fit reverberation mapping estimate from Peterson et al.\ (2004), $5.4 \times 10^6 \Msun$,
is consistent with this estimate.

We find $r$ = $6.0^{+10.2}_{-2.4} \times 10^{13}$ m, or $2.3^{+3.9}_{-1.0}$ lt-days.
As 1 $R_{\rm g}$ corresponds to $1 \times 10^{10}$ m for the black hole mass used, $r$ = $6000^{+10500}_{-2500} R_{\rm g}$.
We cannot of course rule out contribution from an even more narrow Gaussian component originating in
even more distant material. 
In the 2002 {\it XMM-Newton} observation, the corresponding measured line width 
(we use our best-fit value of $\sigma = 87 \pm 17$ eV) corresponds to a value of $r$ = 
$1.3^{+0.7}_{-0.4} \times 10^{13}$ m = $0.50^{+0.30}_{-0.15}$ lt-days, or $1350^{+750}_{-350} R_{\rm g}$ (see also B07).
B07 also used the lack of observed variability in the Fe line flux during the {\it XMM-Newton} observation to constrain the
light-crossing time for the line-emitting gas to be at least 2000 $R_{\rm g}$.

One possible explanation to explain the change in Fe K$\alpha$ line profile,
insofar as it traces the geometrically thin, radiatively efficient disk, is that the innermost radius
of the thin disk has increased over 5 years. A common model for accretion flows incorporating truncated thin disks
is one where the thin disk transitions to an inner RIAF as the flow 
crosses a certain transition radius $r_{\rm t}$ (Esin et al.\ 1997);
a commonly invoked type of RIAF is an advected-dominated accretion flow (ADAF), wherein the disk 
is optically thin and geometrically thick (e.g., Narayan \& Yi 1995). 
The largest width observed for the Fe K$\alpha$ line thus could indicate $r_{\rm t}$.
In this model, $r_{\rm t}$ is expected to increase, and more of the inner thin disk evaporates, as the accretion rate relative 
to Eddington, $\dot{m}$, decreases in a given object.
Supporting evidence for this comes from timing observations of 
black hole X-ray Binary systems during outburst decay:
characteristic temporal frequencies in the power spectral density function (PSD), such as peaks of Lorentzian
components and/or quasi-periodic oscillations, migrate towards lower temporal frequencies as $\dot{m}$ decreases
and the source luminosity fades, as the source evolves through the low/hard spectral state into quiescence
(e.g., Axelsson et al.\ 2005, Belloni et al.\ 2005).  In addition,
the temperature and flux of the soft, thermally emitted component have been seen to decrease with $\dot{m}$
in many sources (e.g., Gierli\'{n}ski, Done \& Page 2008).

However, the predicted relationship between $\dot{m}$ and 
$r_{\rm t}$ remains unclear. Yuan \& Narayan (2004) empirically derive that
compact sources accreting at $\dot{m}$ near $10^{\sim-2}$,  $10^{\sim-4}$, and  $10^{\sim-(6-7)}$
may be associated with values of $r_{\rm t}$ near  $10^{\sim(1-2)}$, $10^{\sim(2-3)}$,  and $10^{\sim(4-5)}$    $R_{\rm g}$, 
respectively. Assuming a 2--10 keV flux in 2002
of $3.9 \times 10^{-11}$ erg cm$^{-2}$ s$^{-1}$ from
{\it RXTE-PCA} monitoring, a luminosity distance of 41.3 Mpc (following Mould et al.\ 2000, and using
$H_{\rm o}$ = 70 km s$^{-1}$ Mpc$^{-1}$ and $\Lambda_{\rm o}$ = 0.73), the 2--10 keV
luminosity $L_{2-10} = 7 \times 10^{42}$ erg s$^{-1}$. From Marconi et al.\ (2004),
an AGN with this $L_{2-10}$ has a 
bolometric luminosity $L_{\rm bol} = 15 L_{2-10} = 1.1 \times 10^{44}$ erg s$^{-1}$.
The accretion relative to Eddington $\dot{m}$ = $L_{\rm bol}/L_{\rm Edd}$ is thus estimated to be 
0.15 for the 2002 {\it XMM-Newton} observation. 
$\dot{m}$ during the {\it Suzaku} observation is thus 0.04. 
Meanwhile, Lu \& Wang (2000) have derived $\dot{m} \sim 5.5\%$ from SED fitting.
These values of $\dot{m}$ and our derived value of $r_{\rm t}$
are not immediately consistent with the rough relation of Yuan \& Narayan (2004), thus pointing toward
models incorporating smaller values of $r_{\rm t}$ (see below).
Furthermore, most low-$\dot{m}$ sources are radio loud, but
NGC 4593 is not strongly radio-loud. 
Its 5 GHz flux has been measured to near $\sim$2 mJy (e.g.,
Schmitt et al. 2001), and its B-band flux is $\sim$6--16 mJy
(e.g., McAlary et al.\ 1983), so the radio loudness parameter,
defined as the ratio of these two values, is $\lesssim$3.
Values $\geq$10 define a source as radio-loud  (Kellermann et al.\ 1989).
A connection between $\dot{m}$ (proportional to the observed X-ray flux) and $r_{\rm t}$ in NGC 4593 
is thus qualitative only as well as speculative, especially since we have only two model-dependent
estimates of $r_{\rm t}$. 

There is also the question of whether the inner portions of a thin disk in AGN can  
evaporate and/or become radiatively inefficient on timescales of only a few years. 
As the accretion disks of BH XRBs are thought to evolve on timescales of at least hours to days,
the corresponding timescales in NGC 4593 (black hole mass a factor of $10^{\gtrsim5}$ higher)
would be decades to centuries. On the other hand, Marscher et al.\ (2002) interpreted 
rapid (duration of $\sim$ a couple weeks) dips 
in the X-ray light curve of the radio-loud Seyfert 3C~120 
as periods when the inner portion of the disk evaporated, each event leading to
ejection of material into the relativistic jet and a corresponding radio flare about a month later.
 


The total Fe K$\alpha$ line intensity decreased from 
$4.79 \pm 0.48 \times 10^{-5}$ (2002) to $2.93 \pm 0.23 \times 10^{-5}$ (2007) ph cm$^{-2}$ s$^{-1}$, a factor of 1.7,
while the observed 4--10 keV flux decreased from $2.47\times 10^{-11}$ (2002) 
to $0.69 \times 10^{-11}$ (2007) erg cm$^{-2}$ s$^{-1}$, a factor of 3.6, i.e., 
roughly half of the total line flux has responded to continuum decrease.
Modeling the Fe K$\alpha$ line with a dual-Gaussian model attempted to separate the 
variable and non-variable emission components; in this model, a non-variable, narrow component
is detected in both observations, and dominates the total line flux in the {\it Suzaku}
spectrum, while a broader component is detected only in the {\it XMM-Newton}
spectrum.  The best-fit width $\sigma$ of the broad line was 
$177^{+84}_{-52}$ eV, corresponding to FWHM velocity of $19100^{+9100}_{-5600}$ km s$^{-1}$,
implying $r = 3.2^{+3.3}_{-1.7} \times 10^{12}$ m = $0.12^{+0.12}_{-0.06}$ lt-days, or 
$330^{+330}_{-180} R_{\rm g}$. This estimate is
inconsistent with B07's estimate of $\sim$2000 $R_{\rm g}$ based on the invariance of the Fe K$\alpha$ line
during the {\it XMM-Newton} observation, but it is 
still consistent with the presence of a truncated thin disk ($r_{\rm t} > 150 R_{\rm g}$).
$r_{\rm t}$ could be invariant from 2002 to 2007, 
but an annular region on the thin disk spanning from $\sim 300 R_{\rm g}$ to 1000--5000 $R_{\rm g}$
(outer radius obviously speculative) could be responding to the drop in illuminating continuum flux.
If the drop in \ion{Fe}{26} flux is real, then that line could also originate in this region.
However, it is not clear in this model why material at $\gtrsim 6000 R_{\rm g}$ (yielding the narrow line component), 
contributing roughly half of the total line intensity in 2002, has not responded to
the drop in continuum flux, as it is well within a week's light-travel time.
The width of the narrow line had been fixed at $\sigma=41$ eV in our modeling of the
{\it XMM-Newton} profile, but contributions from a more narrow component 
likely cannot be ruled out. Such distant material could be responding to a previous higher 
continuum flux. The 2--10 keV {\it RXTE}-PCA monitoring light curve in fact
showed a higher, more average flux level until roughly
300 days before the {\it Suzaku} observation (Figure 7).

A final possibility that does not require evolution in $r_{\rm t}$
is that the inner disk may have become be too ionized to transmit an Fe line.
In the context of models with a hot, ionized skin 
(Nayakshin, Kazanas \& Kallman 2000),
a disk illuminated by a power-law continuum with a photon index near 1.6, similar to that in the
{\it Suzaku} observation, yields extremely weak Fe line emission.

Of course, both profile models are likely oversimplifications.
The community could thus benefit from an X-ray observatory with $\sim$ few eV resolution 
combined with a large effective area near 6 keV to resolve the various 
components of the Fe K$\alpha$ core as a function of time
and/or continuum flux level, if multiple components do indeed exist,
as well as resolve the \ion{Fe}{26} line.


\subsection{The Compton Reflection Component}    

A Compton shoulder was not significantly detected in either the {\it Suzaku} or {\it XMM-Newton} spectra;
we find upper limits to Compton shoulder emission (first-scattering) of
$\sim$23$\%$ of the core. It is thus not obvious from this limit 
alone whether bulk of the Fe K$\alpha$ line originates in Compton-thick material, especially
the degree to which the strength of the Compton shoulder depends on the geometry of the material.
However, no relativistically broadened Fe K$\alpha$ line has been confirmed
in NGC 4593, but we confirm from the broadband {\it Suzaku} spectrum the presence of a Compton reflection hump
which thus must correspond to (at least some fraction of) the Fe K$\alpha$ core emission.
We can investigate if the measured strength $R$ of the Compton reflection hump, 
$1.08 \pm 0.20$ (statistical uncertainty only; $\pm$ 0.35 including the systematic uncertainty),
can correspond to the observed Fe K$\alpha$ line $EW$ of 255$\pm$19 eV. 
Following George \& Fabian (1991), one expects an $EW$ of 135 eV (using the abundances of Lodders 2003)
to correspond to $R$ = 1 
for an semi-infinite optically thick slab illuminated by a power-law continuum with $\Gamma$ = 1.7
and assuming solar abundances and an inclination angle of 30$\degr$ relative to
the observer's line of sight. 
The observed values of $R$ and $EW$ are thus
consistent if the Fe abundance relative to solar, $Z_{\rm Fe}$, is about 1.7, which is not unreasonable. 
For a truncated disk, this could be explained by having the 
Comptonizing corona consist of numerous flares lying in a sandwich-like geometry just above/below the thin disk
(e.g., Haardt et al.\ 1994), 
such that the disk spans 2$\pi$ sr of the sky as seen by each X-ray flare.
However, very distant (pc-scale), Compton-thick material lying out of the line of sight, which cannot be ruled out
as contributing to the observed Fe K emission profile, could also contribute to the total
Compton reflection strength.
   
However, if the thin disk is truncated, then a semi-infinite slab may not be
an appropriate geometry, especially if the
central X-ray source is not immediately close to the reflecting disk. 
The $EW$ of a truncated disk will be lower, but will depend on the location of the illuminating X-ray source. 
If we assume that the illuminating X-ray source is located on the disk symmetry axis a height $h$
above the disk, we can use the $EW$ to constrain $h$. George \& Fabian (1991, their Fig.\ 15)  
demonstrate that the reflected flux is dominated by the region of the disk with $r/h \sim$ 1--2.
For a truncated thin disk with $r_{\rm t}$ = several thousand $R_{\rm g}$, 
$h$ must also be $\sim$ several thousand $R_{\rm g}$ above the black hole. A configuration
in which the X-ray corona is associated with the base of a jet along the symmetry axis
may thus be applicable, e.g., Markoff, Nowak \& Wilms (2005).
NGC 4593, like many Seyferts, is known to host a pc-scale radio component
(size of $<$15 pc; Schmitt et al.\ 2001).





\subsection{Spectral Variability of the Broadband Components}


The primary-law component in Seyfert X-ray spectra
is usually attributed to inverse Comptonization of soft seed photons.
In an ADAF flow, thermal Comptonization is expected to dominate the X-ray emission 
unless the accretion rate is extremely low, in which case thermal
bremsstrahlung emission dominates the X-ray spectrum (Narayan et al.\ 1998, Narayan 2005). 
One of our main results is that while the primary power-law component has dropped
in flux by a factor of almost 4, the soft excess has dropped in flux by a factor of $\sim$20 
between 2002 and 2007, ruling out an origin for the soft excess with a size greater than 5 lt-years.
This drop is likely linked to the decrease in the primary X-ray power-law, i.e., it may be
either a cause of an effect of it.

We explored two phenomenological models for the soft excess, bremsstrahlung and thermal Comptonization.
In the presence of an ADAF flow, 
one can expect thermal bremsstrahlung emission with a temperature of $10^{9-11}$ K
(Narayan \& Yi 1995), but such temperatures are higher by over 2 orders of magnitude compared to
the temperature derived from our model fits and by B07.
We also modeled the soft excess as inverse Comptonization of soft seed photons with an assumed
input temperature of 50 eV by an optically thin corona with a temperature $k_{\rm B}T \sim 50$ keV, 
again obtaining similar results to B07. The location of such a process is not clear, though it could occur 
in the ionized skin of the thin disk (e.g., Magdziarz et al.\ 1998, Janiuk et al.\ 2001), 
or at the base of an outflowing jet. If both the soft excess and hard X-ray power-law components
originate via Comptonization of disk seed photons, a decrease in the intensity of those soft seed
photons (e.g., from an increase in $r_{\rm t}$) between 2002 and 2007
could yield a correlated drop in both component's flux.
Another possibility is that the optical depth of the Comptonizing components may have changed.



\acknowledgements This research has made use of data obtained from the {\it Suzaku} 
satellite, a collaborative mission between the space agencies of Japan (JAXA) and the USA (NASA).
This work has made use of HEASARC online services, supported by
NASA/GSFC, and the NASA/IPAC Extragalactic Database,
operated by JPL/California Institute of Technology under
contract with NASA. A.M.\ acknowledges financial support from NASA grant
NNX08AL36G.











\appendix
In this Appendix, we discuss the X-ray spectrum of 1WGA J1239.7--0526,
an X-ray source located about 5$\arcmin$ south of NGC 4593,  
at $\alpha$ = 12h39m42.8s, $\delta$ = --5d26m15s (J2000.0 coordinates from the
NASA Extragalactic Database), and in the {\it Suzaku}-XIS field of view during 
the observation of NGC 4593.
Detection of this source in X-rays was first reported by {\it ROSAT}-PSPC (White, Giommi \& Angelini 2000).
Its Galactic latitude is +57$\degr$, so it is likely not Galactic in nature.
Its redshift is unknown.
As this source is also in the field of view of the (non-imaging) HXD-PIN
observation of NGC 4593, we extracted spectra of 1WGA J1239.7--0526 from the XIS to
estimate the level of contamination in the HXD-PIN spectrum of NGC 4593.

Spectra were extracted in a manner similar to Section 2.1. 
ARFs were generated considering the source's position 5$\arcmin$ off-axis.
Spectra were grouped to a minimum of 20 ct bin$^{-1}$.

We fit the co-added 0.4--11 keV XIS-FI and 0.3--9 keV XIS-BI spectra in XSPEC
with a model consisting of a simple power-law, with absorption by the
Galactic column (1.88 $\times$ 10$^{20}$ cm$^{-2}$, Kalberla et al.\ 2005).
This model yields a good fit with excellent data/model residuals, with $\xd$ = 369.0/378
for $\Gamma = 1.75 \pm 0.05$ and a power-law normalization of 
$ 1.05 \pm 0.05 \times 10^{-4}$ ph cm$^{-2}$ s$^{-1}$ keV$^{-1}$ at 1 keV. 
Absorbed fluxes for the 0.4--2 and 2--10 keV bands are
$2.4 \times 10^{-13}$ and $3.9 \times 10^{-13}$ erg cm$^{-2}$ s$^{-1}$, respectively.
Background fluxes were 0.31 (FI) and 0.53 (BI) times
the modeled source flux in the 0.4--2.0 keV band,
and 0.58 (FI) and 2.2 (BI)
the modeled source flux in the 2--10 keV band.

Extrapolating this model to the HXD-PIN band yields a 12--76 keV flux estimate of
$ 7.3 \times 10^{-13}$ erg cm$^{-2}$ s$^{-1}$, or about 1.5\% of the net source flux of NGC 4593
in the 12-76 keV band. 
However, the HXD aimpoint is about 4$\arcmin$ away from 1WGA J1239.7--0526; at this angular distance,
the PIN effective area is about 90$\%$ of that on-axis;
1WGA J1239.7--0526 thus contributes only $\sim$1.35\% of the net count rate
of NGC 4593 in the PIN.

To test for the presence of variability in 1WGA J1239.7--0526,
we extracted light curves from each XIS, binned them to 34560 s (6 satellite orbits),
and added light curves from all three XISes.
In the 0.4--2.0 keV band, an upper limit to $F_{\rm var}$ of 15.2$\%$ is found.
In the 2--10 keV band, we measured the variability amplitude $F_{\rm var} = 9\pm6\%$.

\begin{deluxetable*}{llcc}\tablecolumns{4}
\tabletypesize{\footnotesize}
\tablewidth{0pc}
\tablecaption{Fe K Bandpass Model Parameters\label{tab1}}
\tablehead{
\colhead{}          & \colhead{}          & \colhead{2007 {\it Suzaku}} & \colhead{2002 {\it XMM-Newton}} \\
\colhead{Component} & \colhead{Parameter} & \colhead{XIS-FI value}      & \colhead{EPIC-pn value}}
\startdata
                            & $\xd$                                  & 320.6/318                  &  1105.0/1091           \\   
Model Flux                  & 4--10 keV flux (erg cm$^{-2}$ s$^{-1}$) &   $0.69 \times 10^{-11}$     & $2.47 \times 10^{-11}$ \\ 
Power-law                   & $\Gamma$                               & $1.647 \pm 0.043$    & $1.839^{+0.016}_{-0.019}$ \\
Compton Reflection          & $R$                                    & 1.08 (fixed)    &  1.08 (fixed) \\
Fe K$\alpha$ emission line  & Energy (keV)                           & 6.412$\pm$0.007                &  6.396 $\pm$ 0.012 \\
                            & Width $\sigma$ (eV)                    & $41^{+12}_{-16}$                & 87 $\pm$ 17 \\
                            & Intensity (ph cm$^{-2}$ s$^{-1}$)       & $2.93 \pm 0.23 \times 10^{-5}$ & $ 4.79 \pm 0.48 \times 10^{-5}$ \\ 
                            & $EW$ (eV)                              &  255$\pm$19                     & 114 $\pm$ 11 \\
\ion{Fe}{25} emission       & Energy (keV)                           & 6.70 (fixed)              & 6.70 (fixed) \\
                            & Width $\sigma$ (eV)                    & 1 (fixed)                 & 1 (fixed)   \\
                            & Intensity (ph cm$^{-2}$ s$^{-1}$)       & $ < 2.8 \times 10^{-6}$  & $ < 5.5 \times 10^{-6}$ \\
                            & $EW$ (eV)                              & $ < 19 $             & $ < 12 $ \\
\ion{Fe}{26} emission line  & Energy (keV)                           & 6.96 (fixed)              & 6.96 (fixed) \\
                            & Width $\sigma$ (eV)                    & 1 (fixed)                & 1 (fixed)    \\
                            & Intensity (ph cm$^{-2}$ s$^{-1}$)       & $< 1.8 \times 10^{-6}$    & $4.4\pm 2.8 \times 10^{-6}$ \\
                            & $EW$ (eV)                              & $<$ 17                   &  12 $\pm$ 8 \\
                            & $F$-test significance                  &   N/A                           &  92$\%$ (97.5$\%$)   \\
Compton Shoulder emission   & Intensity (ph cm$^{-2}$ s$^{-1}$)       & $< 7 \times 10^{-6}$    & $< 1.1 \times 10^{-5}$ \\
                            & $EW$ (eV)                              & $<$ 44                   &  $ < 22 $ \\
\enddata
\tablecomments{ 
Results are shown for the respective best-fit models for the
4--11.5 keV 2007 {\it Suzaku} XIS-FI spectrum and for the 4--11.5 keV 2002 {\it XMM-Newton} pn spectrum in which the
Fe K$\alpha$ line is modeled as a single Gaussian.
In each case, the best-fit model consists of a power-law component, a Compton reflection component ({\sc pexrav}), and
emission lines due to \ion{Fe}{1} K$\alpha$ and \ion{Fe}{1} K$\beta$; a \ion{Fe}{26} emission line was
included in the fit to the {\it XMM-Newton} pn data.
In the {\it Suzaku} data, 
the Fe K$\alpha$ line energy has been corrected to account for the gain shift of +9 eV as measured using the $^{55}$Fe 
calibration source spectra.
The row labelled ``$F$-test significance'' denotes the confidence level associated with including this component in the model
according to an $F$-test (omitted where line intensity is an upper limit only); values in parentheses denote 
statistical significance derived from Monte Carlo simulations (see Section 3 for details).}
\end{deluxetable*}

\begin{deluxetable*}{llc}\tablecolumns{3}
\tabletypesize{\footnotesize}
\tablewidth{0pc}
\tablecaption{{\it Suzaku} Broadband Spectral Fit Parameters\label{tab2}}    
\tablehead{
\colhead{Component} & \colhead{Parameter} & \colhead{Value}  }
\startdata
                            & $\xd$                                  & 1254.5/1106                       \\   
Model Flux                  & 2--10 keV flux (erg cm$^{-2}$ s$^{-1}$) &   $1.03 \times 10^{-11}$      \\ 
Power-law                   & $\Gamma$                               & $1.656^{+0.018}_{-0.025}$   \\
                            & Norm.\ at 1 keV (ph cm$^{-2}$ s$^{-1}$ keV$^{-1}$) & $2.08 \pm 0.04 \times 10^{-3}$ \\
Warm Absorber               & Column density $N_{\rm H,WA}$ (cm$^{-2}$) & $ 2.8 \pm 0.8 \times 10^{21}$ \\
                            & log($\xi$) (erg cm s$^{-1}$)             & 2.39 $\pm$ 0.17 \\
Compton Reflection          & Refl.\ strength $R$                     & 1.08 $\pm$ 0.20 \\
Bremss.\ Component           & $k_{\rm B}T$ (keV)                        & $0.20^{+0.06}_{-0.05}$ \\       
                            & Norm.\                                  & $0.8^{+1.9}_{-0.5} \times 10^{-3}$ \\ 
                            & Unabsorbed (absorbed) $F_{\rm 0.4-2.0 keV}$  & $8.7^{+12.2}_{-6.1}$ ($7.1^{+9.9}_{-5.0}$) $\times 10^{-14}$ \\ 
Fe L$_3$ Edge                  & Energy (keV)                           & 0.707 (fixed) \\
                            & $\tau$                                  & $0.09^{+0.03}_{-0.04}$ \\
                            & $F$-test significance                   & 99.7$\%$ (96.4$\%$) \\       
\ion{O}{7} $f$ Emission Line    & Energy (keV)                           & 0.561 (fixed) \\                                   
                            & Intensity (ph cm$^{-2}$ s$^{-1}$)       & $2.9 \pm 2.8 \times 10^{-5}$ \\ 
                            & $EW$ (eV)                              & $4.9 \pm 4.7$ \\ 
                            & $F$-test significance                   & 90$\%$ (84$\%$)\\       
\ion{Ne}{9} $f$ Emission Line    & Energy (keV)                       & 0.905 (fixed) \\                                   
                            & Intensity (ph cm$^{-2}$ s$^{-1}$)       & $1.1^{+0.6}_{-0.7} \times 10^{-5}$ \\ 
                            & $EW$ (eV)                              & $4.5^{+2.3}_{-2.7}$ \\ 
                            & $F$-test significance                   & 98.4$\%$ (95$\%$) \\       
 \ion{Mg}{11} $f$ Emission Line    & Energy (keV)                     & 1.331 (fixed) \\                                   
                            & Intensity (ph cm$^{-2}$ s$^{-1}$)       & $4.0^{+3.1}_{-1.6} \times 10^{-6}$ \\ 
                            & $EW$ (eV)                              & $3.0^{+2.3}_{-1.2}$ \\ 
                            & $F$-test significance                   & 98.9$\%$ (94$\%$) \\       
XIS-BI/XIS-FI               & Norm.\                                 & $1.059^{+0.010}_{-0.007}$ \\
\enddata
\tablecomments{Best-fit model parameters  for the {\it Suzaku} broadband fit with the soft excess modeled
as thermal bremsstrahlung emission. See the {\sc xspec} user manual for the units of the bremsstrahlung
normalization. 
The rows labeled ``$F$-test significance'' denote the confidence level associated with including this component in the model
according to an $F$-test; values in parentheses denote 
statistical significance derived from Monte Carlo simulations (see Section 3 for details). 
The uncertainty on $R$ is statistical only; the systematic uncertainty is an additional $\pm$0.15.}
\end{deluxetable*}

\begin{deluxetable*}{llcc}\tablecolumns{4}
\tabletypesize{\footnotesize}
\tablewidth{0pc}
\tablecaption{Comparison of Broadband Model Fits to {\it XMM-Newton} and {\it Suzaku} spectra\label{tab3}}
\tablehead{
\colhead{}          & \colhead{}          & \colhead{2002 {\it XMM-Newton}} & \colhead{2007 {\it Suzaku}}  \\
\colhead{Component} & \colhead{Parameter} & \colhead{EPIC-pn value}     &   \colhead{value}      }
\startdata
\multicolumn{4}{c}{Thermal Bremsstrahlung model} \\ \hline      
                            & $\xd$                                      & 1793.23/1756  & 1254.8/1105 \\          
(Absorbed) Model Flux       & 2--10 keV flux (erg cm$^{-2}$ s$^{-1}$)     & $3.93 \times 10^{-11}$ & $1.03 \times 10^{-11}$ \\
                            & 0.5--2.0 keV flux (erg cm$^{-2}$ s$^{-1}$)  & $2.57 \times 10^{-11}$ & $4.15 \times 10^{-12}$\\
Power-law                   & $\Gamma$                                   & 1.901$\pm$0.006  & $1.656^{+0.018}_{-0.025}$ \\
                            & Norm.\ at 1 keV (ph cm$^{-2}$ s$^{-1}$ keV$^{-1}$) & $1.19 \pm 0.01 \times 10^{-3}$  & $2.08 \pm 0.04 \times 10^{-3}$ \\
Warm Absorber               & Column density $N_{\rm H,WA}$ (cm$^{-2}$) & $ 1.7 \pm 0.2  \times 10^{21}$  & $ 2.8 \pm 0.8 \times 10^{21}$ \\
Fe L$_3$ edge                  & Optical depth $\tau_{\rm FeL3}$           & $0.16 \pm 0.2 $  &  $0.09^{+0.03}_{-0.04}$ \\
Bremss.\ Component           & $k_{\rm B}T$ (keV)                       & 0.28 $\pm$ 0.01  & $0.20^{+0.06}_{-0.05}$ \\       
                            & Norm.\                                  & $2.02 \pm 0.01 \times 10^{-2}$  & $0.8^{+1.9}_{-0.5} \times 10^{-3}$ \\ 
                            & Unabsorbed (absorbed) $F_{\rm 0.4-2.0 keV}$  & $4.72 \pm 0.19$ ($3.87 \pm 0.15$) $\times 10^{-12}$  &  $8.7^{+12.2}_{-6.1}$ ($7.1^{+9.9}_{-5.0}$) $\times 10^{-14}$\\  \hline
\multicolumn{4}{c}{Thermal Comptonization model} \\ \hline     
                            & $\xd$                                      & 1945.8/1757 &  1255.775 / 1105 \\          
(Absorbed) Model Flux       & 0.5--2.0 keV flux (erg cm$^{-2}$ s$^{-1}$)  & $2.57 \times 10^{-11}$  & $4.15 \times 10^{-12}$ \\
Power-law                   & $\Gamma$                                   & $1.750^{+0.018}_{-0.052}$  & $1.659^{+0.13}_{-0.011}$  \\
                            & Norm.\ at 1 keV (ph cm$^{-2}$ s$^{-1}$ keV$^{-1}$) & $8.30^{+0.39}_{-1.10} \times 10^{-3}$ & $2.09^{+0.05}_{-0.04} \times 10^{-3}$ \\
Warm Absorber               & Column density $N_{\rm H,WA}$ (cm$^{-2}$) & $2.3^{+0.3}_{-0.2} \times 10^{21}$  & $3.1^{+1.2}_{-0.9} \times 10^{21}$ \\
Fe L$_3$ edge                  & Optical depth $\tau_{\rm FeL3}$           & $0.15^{+0.01}_{-0.02}$  & 0.08$\pm$0.04\\
Comptonized Component       & Temperature (eV)                 &  53 (fixed)  &  $53^{+57}_{-31}$ \\
                            & Optical depth $\tau$  &  $0.71^{+0.03}_{-0.05}$ &  $<$0.28 \\
                            & Norm.   & $1.60^{+0.17}_{-0.07} \times 10^{-3}$  &  $2.8^{+18.7}_{-2.3} \times 10^{-5}$ \\
                            & Unabsorbed (absorbed) $F_{\rm 0.4-2.0 keV}$  &  $1.54^{+0.33}_{-0.13}$  ($1.28^{+0.27}_{-0.11}$) $\times 10^{-11}$   & $8.2^{+54.6}_{-6.7}$  ($6.7^{+44.4}_{-5.5}$) $\times 10^{-14}$   \\
\enddata
\tablecomments{Results from applying the best-fit broadband models from the {\it Suzaku} observation to
the 0.2--12 keV {\it XMM-Newton} pn spectrum, using thermal bremsstrahlung or thermal Comptonization
components to model the soft excess. Col.\ (2) lists the parameters which were allowed to vary from
their values in the {\it Suzaku} fit; the values for the {\it XMM-Newton} fits are listed in Col.\ (3).
The corresponding values from the {\it Suzaku} fits are listed in Col.\ (4).
See Table 2 for other model parameters for the thermal bremsstrahlung fit. 
Model fluxes are in units of erg cm$^{-2}$ s$^{-1}$. See the {\sc xspec} user manual for the units of the Comptonized component.}
\end{deluxetable*} 

\begin{figure}
\epsscale{0.55}
\plotone{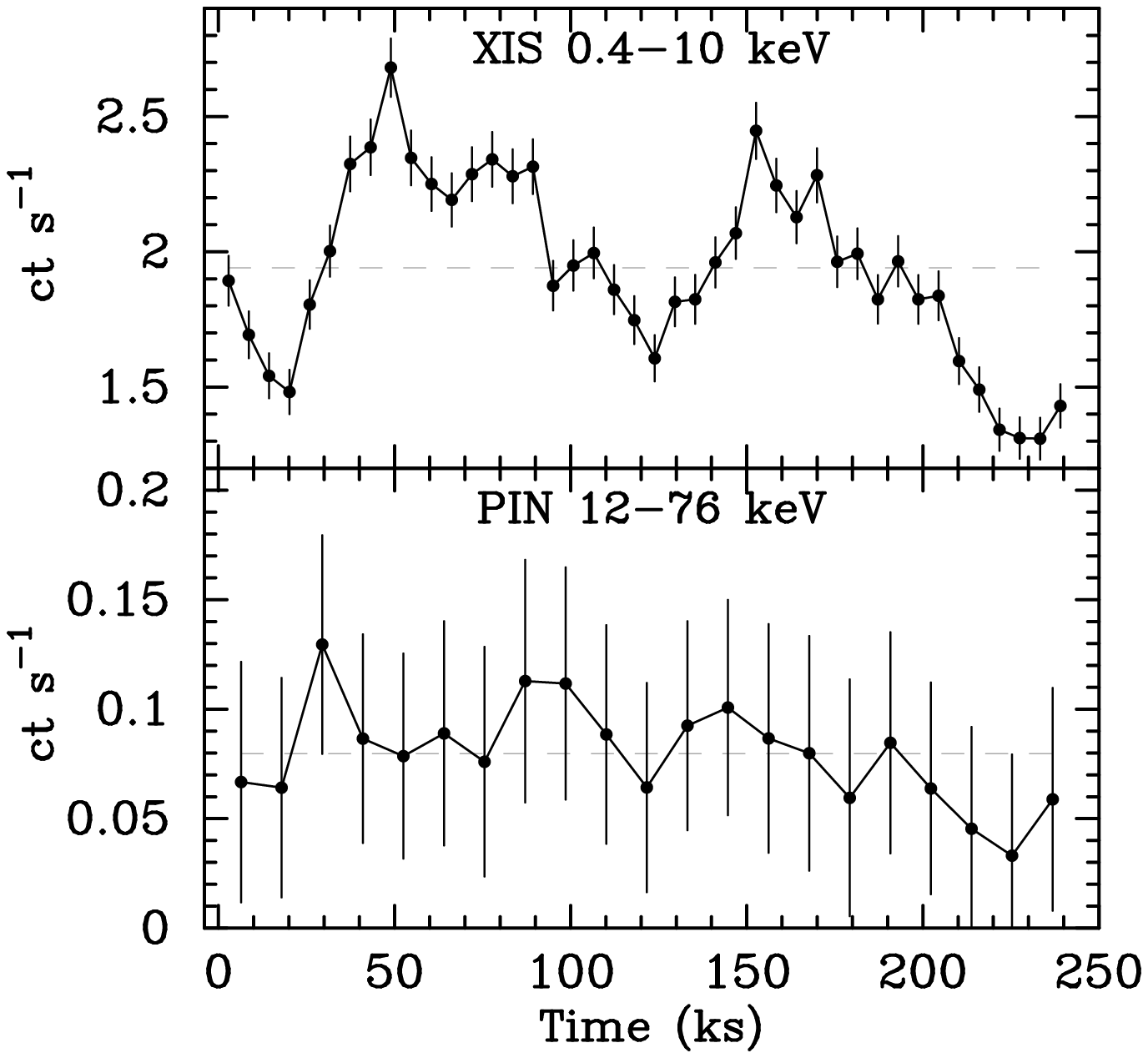}
\caption{Binned light curves from the {\it Suzaku} observation.
The top panel shows the 0.4--10 keV light curve, summed over 
XIS 0+1+3 and binned to 5760 s. The bottom panels shows the
12--76 keV PIN light curve, binned to 11520 s; error bars include a $\sim$2$\%$
systematic uncertainty (Fukazawa et al.\ 2009).
The x-axis denotes time since 2007 December 15 at 04:09 UT.}
\end{figure}

\begin{figure}
\epsscale{0.75}
\plotone{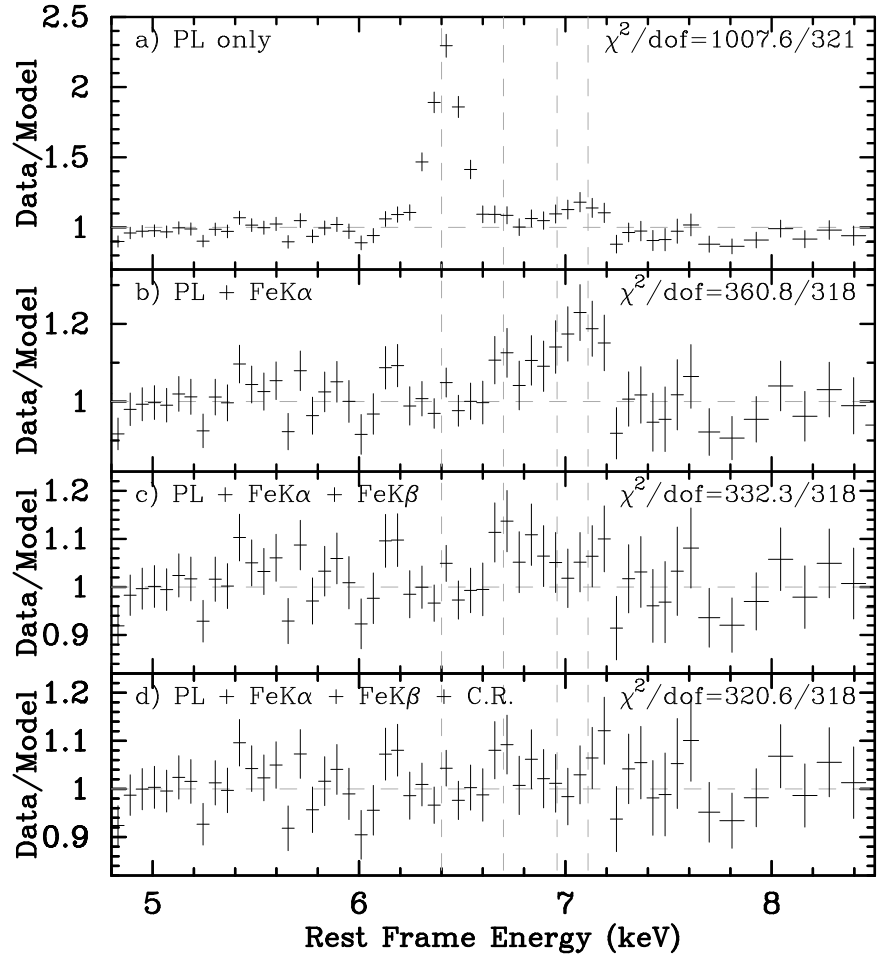}
\caption{Data/model residuals to fits to the {\it Suzaku} XIS-FI Fe K bandpass spectrum.
Data have been rebinned by a factor of 4. 
Residuals to a simple power-law fit are shown in panel (a).
Panel (b) shows residuals when the Fe K$\alpha$ line is modeled with a Gaussian.
In panel (c), a Gaussian to model Fe K$\beta$ line emission has been included.
In panel (d), a Compton reflection component, containing an Fe K edge at 7.11 keV,
has been included; this is our best-fit model.
Vertical dashed lines denote the expected energies of \ion{Fe}{1} K$\alpha$, \ion{Fe}{25},
\ion{Fe}{26}, and \ion{Fe}{1} K$\beta$ emission lines.}
\end{figure}

\begin{figure}
\epsscale{0.50}
\plotone{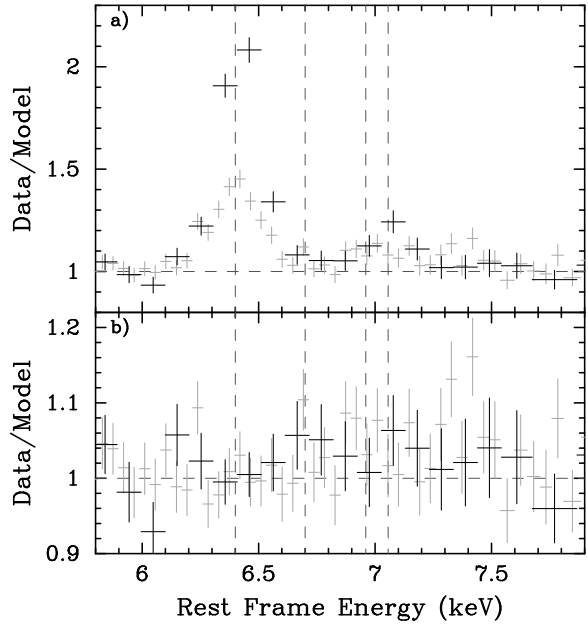}
\caption{Data/model residuals to fits to the {\it Suzaku} XIS-FI 
(black points; rebinned by a factor of 7) and 2002 {\it XMM-Newton} pn
(gray points; rebinned by a factor of 9) Fe K bandpass spectra.
Vertical dashed lines denote the expected energies of Fe K $\alpha$, \ion{Fe}{25}, 
\ion{Fe}{26}, and Fe K$\beta$ emission lines.
Panel (a) shows residuals to a model consisting of a power-law plus a Compton reflection hump;
panel (b) shows residuals when the Fe K$\alpha$ and K$\beta$ lines are included in the model as well.}
\end{figure}

\begin{figure}
\epsscale{0.55}
\plotone{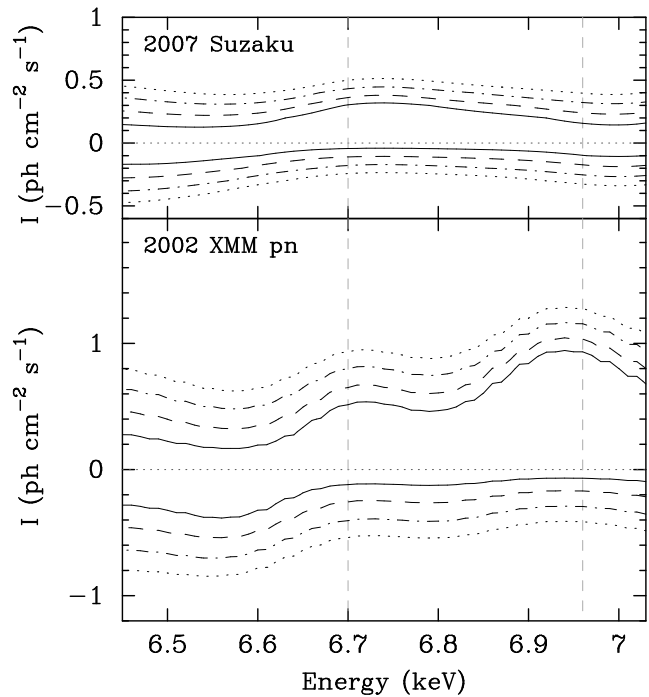}
\caption{Contour plots resulting from applying a ``sliding Gaussian''
to a model consisting of a power-law, a Compton reflection component, and
the Fe K$\alpha$ line.
The resulting plots of line intensity versus energy are shown for the
{\it Suzaku} XIS-FI (upper) and {\it XMM-Newton} pn (lower) Fe K bandpass spectra.
The areas outside the contours indicate regions of (energy, intensity) parameter
space where adding a Gaussian to this model would make the fit worse
The rest-frame energies 
of \ion{Fe}{25} and \ion{Fe}{26} are denoted by vertical dashed lines.
Solid, dashed, dot-dashed, and dotted lines denote 1$\sigma$, 2$\sigma$, 3$\sigma$, and 4$\sigma$
confidence levels, respectively, for two interesting parameters. 
Note that the y-axis scales are the same.
This figure indicates that the \ion{Fe}{26} line detected in the {\it XMM-Newton} spectrum is not
significantly detected in the {\it Suzaku} spectrum.
}
\end{figure}

\begin{figure}
\epsscale{0.75}
\plotone{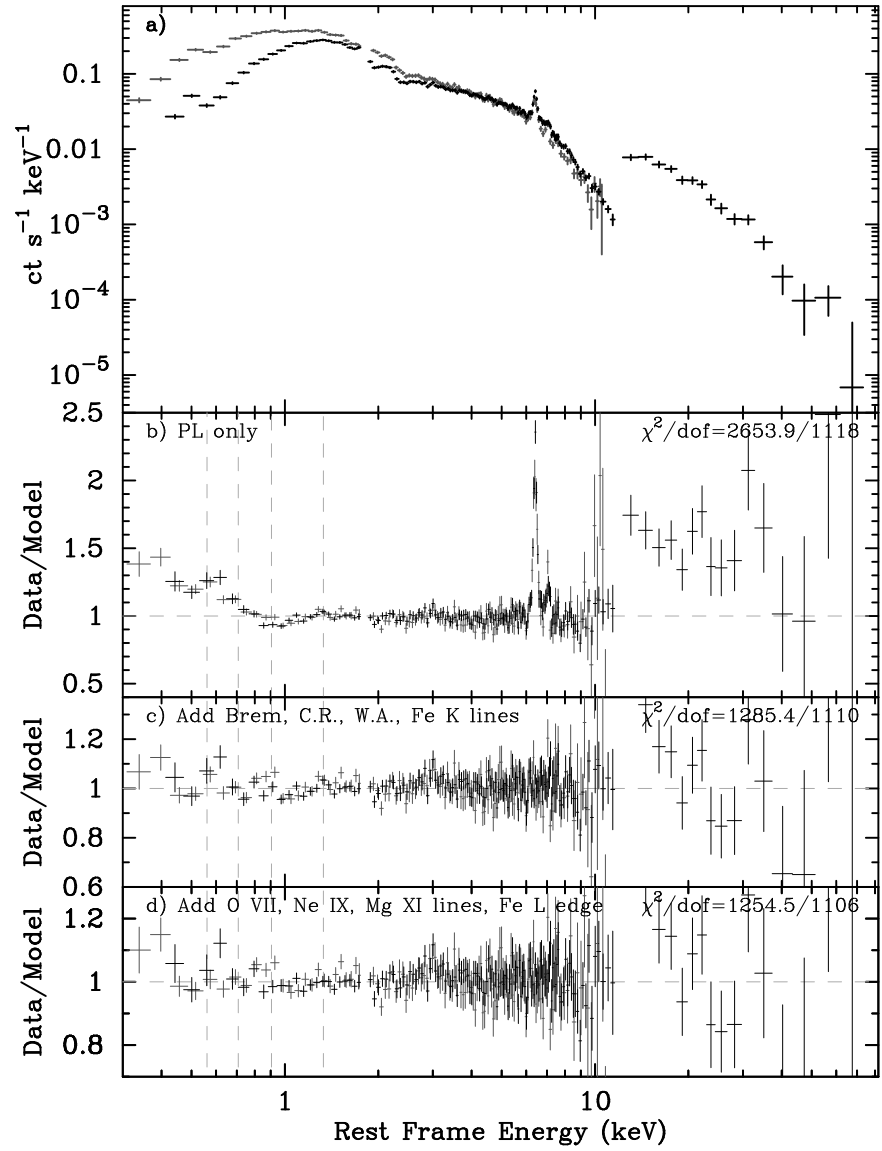}
\caption{Data/model residuals to the broadband spectrum for the {\it Suzaku}
XIS-FI (black points $<$ 12 keV), XIS-BI (grey points), and HXD-PIN (black points $>$ 12 keV). 
The data are plotted in panel (a).
Residuals to a simple power-law fit (panel (b); assuming PIN/XIS-FI normalization of 1.16) 
show the moderately strong
soft excess as well as Fe L absorption features near 0.9--1.2 keV associated with
the warm absorber. The hard excess above 10 keV due to the Compton reflection hump
is also evident.
In panel (c), the Compton reflection component, warm absorber, Fe K emission lines
and soft excess (modeled as thermal bremsstrahlung emission) have been modeled.
In panel (d), the Fe L$_3$ edge and the He-like lines of O, Ne, and Mg have been added.
Vertical dashed lines denote the expected energies of the \ion{O}{7} line, 
the Fe L$_3$ edge energy, the \ion{Ne}{9} line, and the \ion{Mg}{11} line.
All data are rebinned by a factor of 4.}
\end{figure}

\begin{figure}
\epsscale{0.50}
\plotone{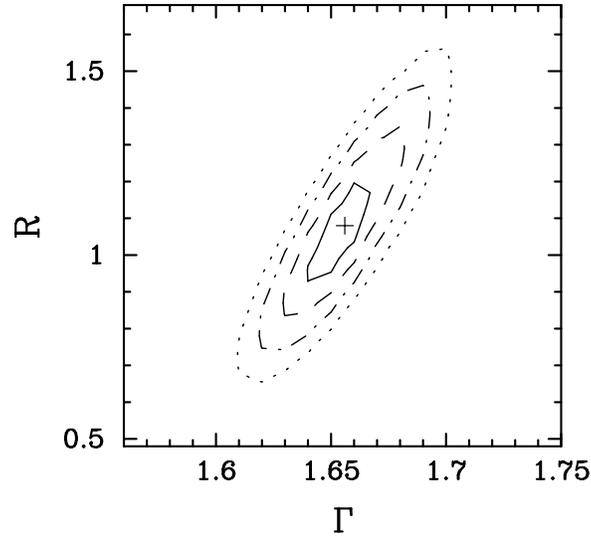}
\caption{Contour plot of Compton reflection component strength $R$ as a function of power-law photon
index $\Gamma$, assuming the PIN/XIS-FI normalization is fixed at 1.16. The cross marks the
best-fit parameters.
Solid, dashed, dot-dashed, and dotted lines denote 1$\sigma$, 2$\sigma$, 3$\sigma$ and 4$\sigma$
confidence levels, respectively, for two interesting parameters.}
\end{figure}

\begin{figure}
\epsscale{0.75}
\plotone{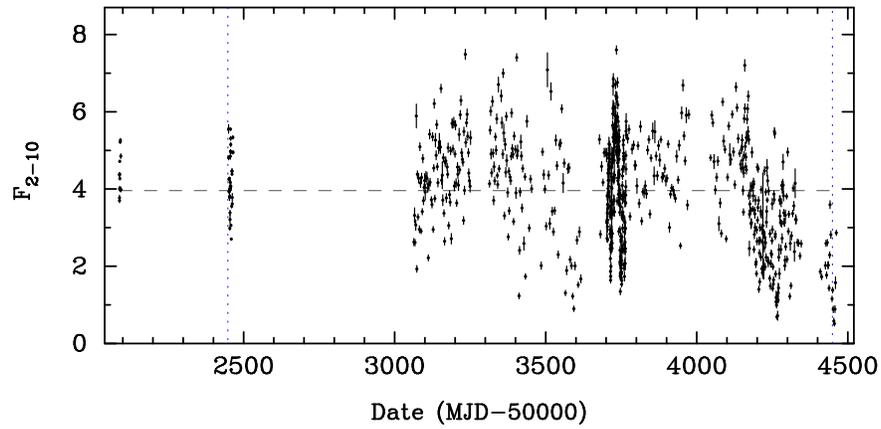}
\caption{2--10 keV light curve from public archive {\it RXTE}-PCA monitoring.
Flux is in units of 10$^{-11}$ erg cm$^{-2}$ s$^{-1}$. The horizontal dashed line denotes
the long-term average mean flux of $3.9 \times 10^{11}$ erg cm$^{-2}$ s$^{-1}$.
Vertical dashed lines denote the times of the 2002 {\it XMM-Newton}
and 2007 {\it Suzaku} observations.}
\end{figure}

\begin{figure}
\epsscale{0.75}
\plotone{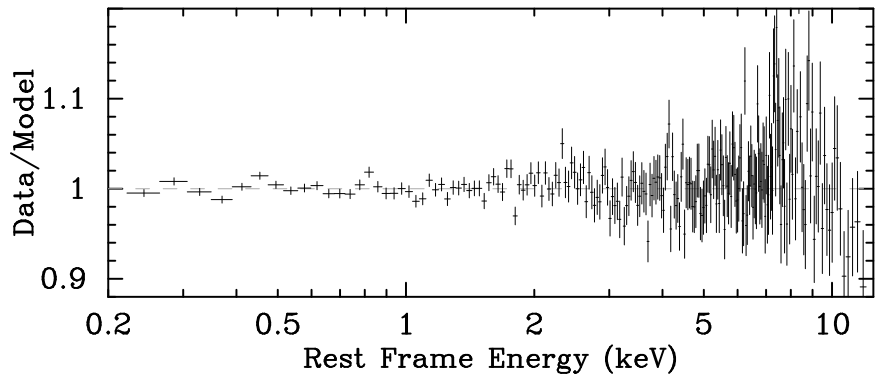}
\caption{Data/model residuals to our best-fit to the 2002 {\it XMM-Newton}
pn spectrum. Data are rebinned by a factor of 8.}
\end{figure}

\begin{figure}
\epsscale{0.4}
\plotone{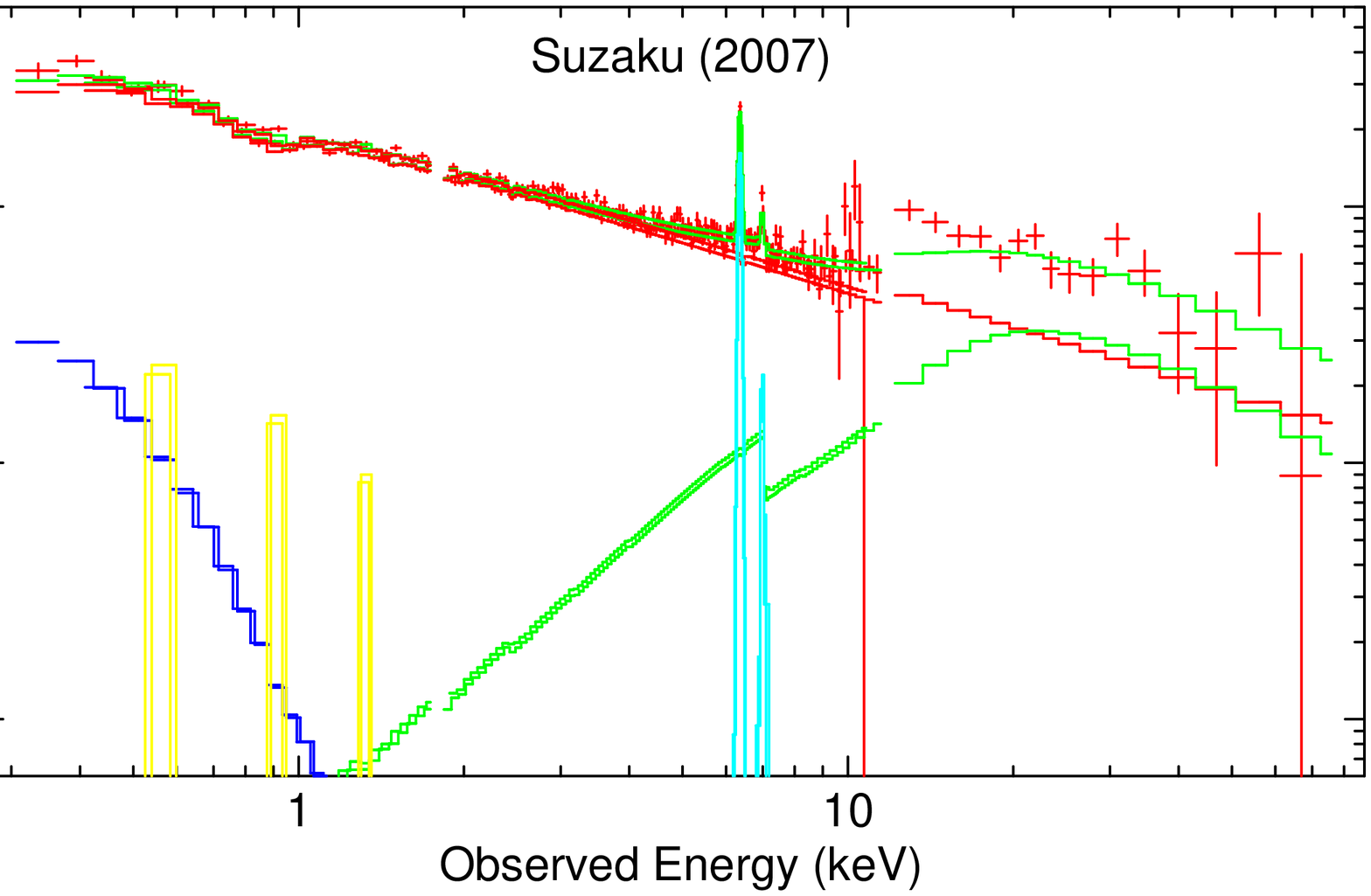}
\caption{Unfolded model spectrum illustrating the best-fit model to
the {\it Suzaku} spectrum. Red crosses denote the data, 
rebinned by a factor of 4. Plotted are the power-law component
(red), Compton reflection component (green), thermal bremsstrahlung component
(blue), He-like O, Ne and Mg lines (yellow), and Fe K$\alpha$ and K$\beta$ 
emission lines (cyan).}
\end{figure}

\begin{figure}
\epsscale{0.4}
\plotone{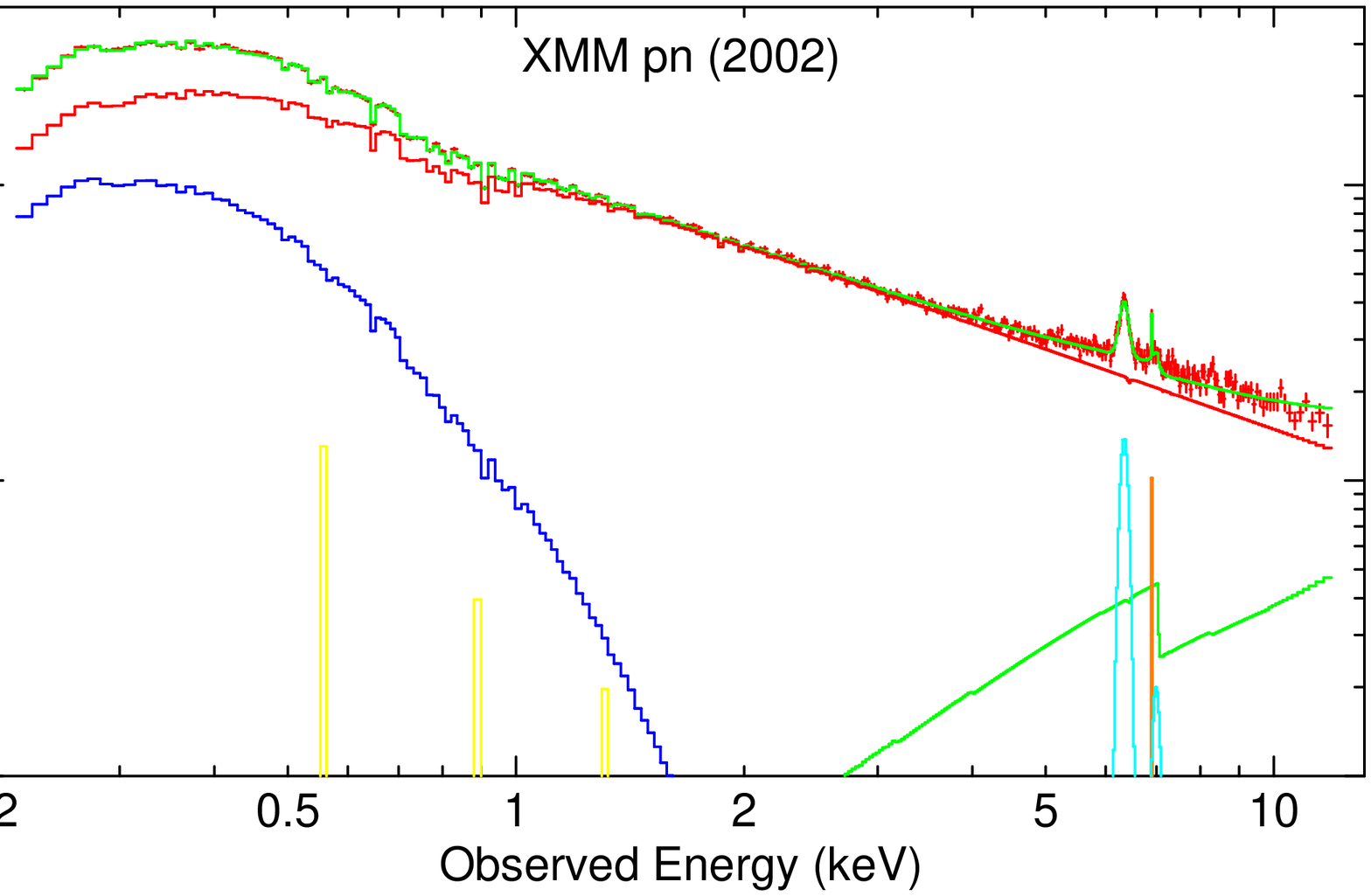}
\caption{Unfolded model spectrum illustrating our best-fit model to
the 2002 {\it XMM-Newton} pn spectrum. Data are binned by a factor of 6.
Components are denoted by the
same colors as in Fig.\ 9; the \ion{Fe}{26} emission line is also plotted in orange.}
\end{figure}


\begin{references}
\reference{A05} Axelsson, M., Borgonovo, L.\ \& Larsson, S. 2005, A\&A, 438, 999
\reference{B05} Belloni, T., Homan, J., Casella, P., et al., 2005, A\&A, 440, 207
\reference{B07} Brenneman, L.W., Reynolds, C.S., Wilms, J.\ \& Kaiser, M.E. 2007, ApJ, 666, 817 (B07)
\reference{EN99} Edelson, R.\ \& Nandra, K. 1999, ApJ, 514, 682
\reference{E97} Esin, A.A., McClintock, J.E.\ \& Narayan, R. 1997, ApJ, 489, 865
\reference{F89} Fabian, A.C., Rees, M.J., Stella, L.\ \& White, N.E. 1989, MNRAS, 238, 729
\reference{F02} Fabian, A.C., Vaughan, S., Nandra, K., et al. 2002, MNRAS, 335, L1
\reference{Fz09} Fukazawa, Y.,  Mizuno, T., Watanabe, S., et al., 2009, PASJ, 61S, 17 
\reference{G86} Gehrels, N. 1986, ApJ, 303, 336
\reference{GF91} George, I.M.\ \& Fabian, A.C. 1991, MNRAS, 249, 352
\reference{G8} Gierli\'{n}ski, M., Done, C.\ \& Page, K. 2008, MNRAS, 388, 753
\reference{Gr99} Gruber, D.E., Matteson, J.L., Peterson, L.E.\ \& Jung, G.V. 1999, ApJ, 520, 124
\reference{G99} Guainazzi, M., Perola, G.C., Matt., G., et al. 1999, A\&A, 346, 607 
\reference{HMG94} Haardt, F., Maraschi, L.\ \& Ghisellini, G. 1994, ApJ,  432, L95 
\reference{IMF07} Ishisaki, Y., Maeda, Y., Fujimoto, R., et al.\ 2007, PASJ, 59S, 113
\reference{J01} Janiuk  A., Czerny  B.\ \& Madejski,  G.M., 2001, ApJ, 557, 408
\reference{K05} Kalberla, P.M.W.\ et al.\ 2005, A\&A, 440, 775 
\reference{Ke89} Kellermann, K. I., Sramek, R., Schmidt, M., Shaffer, D. B., \& Green, R. 1989, ApJ, 98, 1195
\reference{Kb07} Kokubun, M., Makishima, K., Takahashi, T., Murakami, T., Tashiro, M., Fukazawa, Y., Kamae, T., Madejski, G.M.\ et al.\ 2007, PASJ, 59S, 53
\reference{Ky07} Koyama, K., Tsunemi, H., Dotani, T., Bautz, M., Hayashida, K., Tsuru, T., Matsumoto, H., Ogawara, Y.\ et al.\ 2007, PASJ, 59S, 23
\reference{L91} Laor, A.\ 1991, ApJ, 376, 90
\reference{Lo03} Lodders, K.\ 2003, ApJ, 591, 1220
\reference{LW00} Lu, Y.\ \& Wang, T. 2000, ApJ, 537, L103
\reference{Md08} Maeda, Y., Someya, K., Ishida, M., et al., 2008, JX-ISAS-SUZAKU-MEMO-2008-06 
\reference{M98}  Magdziarz,  P., Blaes,  O.M., Zdziarski,  A.A., Johnson, W.N.\ \& Smith  D.A. 1998, MNRAS, 301, 179
\reference{MZ95} Magdziarz, P.\ \& Zdziarski, A.A.\  1995, MNRAS, 273, 837
\reference{Mc04} Marconi, A. et al. 2004, MNRAS, 351, 169
\reference{M05} Markoff, S., Nowak, M.A.\ \& Wilms, J. 2005, ApJ, 635, 1208
\reference{M03} Markowitz, A., Edelson, R., Vaughan, S., et al.\ 2003, ApJ, 593, 96 
\reference{Ma02} Marscher, A.P., Jorstad, S.G., G\'omez, J.-L., et al., 2002, Nature, 417, 625
\reference{Mt02} Matt, G. 2002, MNRAS, 337, 147
\reference{M83} McAlary, C.W., McLaren, R.A., McGonegal, R.J., Maza, J., 1983, ApJS, 52, 341
\reference{McK03} McKernan, B., Yaqoob, T., George, I.M.\ \& Turner, T.J. 2003, ApJ, 593, 142
\reference{Md07} Mitsuda, K., Bautz, M., Inoue, H., Kelley, R., Koyama, K., Kunieda, H., Makishima, K., Ogawara, Y.\ et al.\ 2007, PASJ, 59S, 1
\reference{Md00} Mould, J.R., Huchra, J.P., Freedman, W.L., et al. 2000, ApJ, 529, 786
\reference{MY09} Murphy, K.D.\ \& Yaqoob, T. 2009, MNRAS, 397, 1549
\reference{N06} Nandra, K. 2006, MNRAS, 368, L62
\reference{N5} Narayan, R. 2005, Ap\&SS, 300, 177
\reference{NMQ98} Narayan, R., Mahadevan, R., \& Quataert, E. 1998, in {\it Theory of Black Hole Accretion Disks}, ed. M. A. Abramowicz, G. Bjornsson, \& J. E. Pringle (Cambridge: Cambridge Univ. Press), 148
\reference{NY95} Narayan, R.\ \& Yi, I. 1995, ApJ, 452, 710
\reference{NKK00} Nayakshin, S., Kazanas, D.\ \& Kallman, T.R. 2000, ApJ, 537, 833
\reference{N04} Nelson, C.H., Green, R.F., Bower, G., Gebhardt, K.\ \& Weistrop, D. 2004, ApJ, 615, 652
\reference{N90} Netzer, H. 1990, in Active Galactic Nuclei, ed.\ T.\ J.-L.\ Courvoisier and M.\ Major (Berlin: Springer), 57
\reference{P04} Peterson, B.M., Ferrarese, L., Gilbert, K.M., et al. 2004, ApJ, 613, 682
\reference{P04} Porquet, D., Reeves, J.N., Uttley, P.\ \& Turner, T.J.\ 2004, A\&A, 427, 101
\reference{PT88} Pounds, K.A.\ \& Turner, T.J. 1988, MmSAI, 59, 261
\reference{P02} Protassov, R., van Dyk, D.A., Connors, A., Kashyap, V.L.\ \& Siemiginowska, A. 2002, ApJ, 571, 545
\reference{Qu01} Quataert, E. 2001, in {\it Probing the Physics of Active Galactic Nuclei}, Peterson, B.M., Pogge, R.W., \& Polidan, R.S.\ eds., ASP Conf.\ Proceedings, Vol.\ 224, p.\ 71
\reference{JNR04} Reeves, J.N., Nandra, K., George, I.M., Pounds, K.A., Turner, T.J.\ \& Yaqoob, T. 2004, ApJ, 602, 648 
\reference{R04} Reynolds, C.S., et al. 2004, MNRAS, 352, 205 
\reference{RF05} Ross, R.R. \& Fabian, A.C., 2005, MNRAS, 358, 211
\reference{S01} Schmitt, H.R., Ulvestad, J.S., Antonucci, R.R.J.\ \& Kinney, A.L. 2001, ApJS, 132, 199
\reference{SS73} Shakura, N.I.\ \& Sunyaev, R.A. 1973, A\&A, 24, 337
\reference{St03} Steenbrugge, K., Kaastra, J.S., Blustin, A.J., et al.\ 2003, A\&A, 408, 921
\reference{Ss92} Strauss, M.A., Huchra, J.P., Davis, M., Yahil, A., Fisher, K.B.\ \& Tonry, J. 1992, ApJS, 83, 29
\reference{SM89} Sun, W.-H.\ \& Malkan, M.A. 1989, ApJ, 346, 68
\reference{T07} Takahashi, T., Abe, K., Endo, M., Endo, Y., Ezoe, Y., Fukazawa, Y., Hamaya, M., Hirakuri, S.\ et al.\ 2007, PASJ, 59S, 35
\reference{T94} Titarchuk, L. 1994, ApJ 434, 570
\reference{T05} Turner, T.J., Kraemer, S.B., George, I.M., Reeves, J.N., Bottorff, M.C.\ 2005, ApJ, 618, 155
\reference{V03} Vaughan, S., Edelson, R., Warwick, R.\ \& Uttley, P. 2003, MNRAS, 345, 1271
\reference{W03} Watanabe, S. et al. 2003, ApJ, 597, L37
\reference{WGA00} White, N.E., Giommi. P.\ \& Angelini, L. 2000 VizieR On-line Data Catalog, 9031, 0 
\reference{YP04} Yaqoob, T.\ \& Padmanabhan, U. 2004, ApJ, 604, 63
\reference{YN04} Yuan, F.\ \& Narayan, R. 2004, ApJ, 612, 724
\end{references}
\end{document}